\documentclass[10pt,aps,nofootinbib,superscriptaddress,twocolumn,preprintnumbers,balancelastpage,longbibliography]{revtex4-1}
\usepackage{amsmath,amssymb,amsthm,amsopn}
\usepackage{epsfig}
\usepackage{psfrag}
\usepackage[colorlinks=true,allcolors=blue!65]{hyperref}
\usepackage{multirow}
\usepackage{empheq}
\usepackage{slashed}
\usepackage{graphicx}
\usepackage{url}
\usepackage{subfigure}
\usepackage{textcomp}
\usepackage{bm}
\usepackage{dcolumn}
\usepackage{booktabs}
\usepackage{color,xcolor}

\usepackage{cancel}
\usepackage{braket}
\usepackage[title]{appendix}
\usepackage{lipsum}
\usepackage{mathptmx}
\DeclareMathAlphabet{\mathcal}{OMS}{cmsy}{m}{n}

\def\comment#1{}

\def\beq{\begin{equation}}
\def\eeq{\end{equation}}
\def\bea{\begin{eqnarray}}
\def\eea{\end{eqnarray}}

\newcommand{\dd}{\mathrm{d}}
\begin{document}

\title{The Higgs boson as a self-similar system: Tsallis form factors and the electroweak hierarchy problem}

\author{M. Ahmadvand}
\email{ahmadvand@ipm.ir}
\affiliation{Faculty of Physics, Shahrood University of Technology, P.O. Box 3619995161, Shahrood, Iran}

%\date{\today}

\begin{abstract}
We study the electroweak hierarchy problem for a composite Higgs boson with a statistically self-similar internal structure. Using the spectral representation of the Higgs propagator, we show that this structure can contribute only through momentum-dependent vertex functions. In a superstatistical model of the self-similar structure, a Gamma-distributed size of the constituents turns a Gaussian vertex form factor into the $q$-exponential (Tsallis) form factor $F_q(k_E^2)=[1+(q-1)k_E^2/M^2]^{-1/(q-1)}$, where $M$ is the compositeness scale and $q-1=2/(3N)$ for an effective number $N$ of constituents. These form factors interpolate between those of nonlocal field theories ($q\to1$) and Lee--Wick-type models ($q=2$). For $1<q<2$ every one-loop contribution to the Higgs mass is finite and heavy states decouple. For the benchmarks $q=11/9$, $4/3$ and $5/3$, tuning is negligible ($\Delta\lesssim1$) for $M$ up to about $1$--$2$~TeV, and remains mild ($\Delta \leq 10$) for $M$ up to about $3$--$5$~TeV. The hierarchy problem is thus converted into the requirement of new composite dynamics near the TeV scale, testable at the High-Luminosity Large Hadron Collider through off-shell and high-momentum Higgs processes and searches for new states near $M$.

\vspace*{0.3cm}
\end{abstract}

\maketitle

%=====================================================================
\section{Introduction}\label{sec:intro}
%=====================================================================
The discovery of a Higgs boson with mass $m_h\simeq125$~GeV~\cite{ATLAS:2012yve,CMS:2012qbp,CMS:2013btf} completed the particle content of the Standard Model (SM), yet left the origin of the electroweak scale unexplained. The problem can be stated in terms of a physical cutoff $\Lambda$, a hard momentum cutoff at the scale up to which the SM is assumed to hold. The quadratically divergent part of the one-loop correction to the Higgs mass then has magnitude~\cite{Veltman:1980mj,Kolda:2000wi}
\begin{equation}
|\delta m_h^2|=\frac{3\Lambda^2}{8\pi^2 v^2}\left|m_h^2+2m_W^2+m_Z^2-4m_t^2\right|+\mathcal{O}\!\left(\ln\frac{\Lambda^2}{m_h^2}\right),
\label{eq:veltman}
\end{equation}
where $v\simeq246$~GeV is the vacuum expectation value (vev) of the Higgs field and $m_{W,Z,t}$ are the $W$, $Z$ and top-quark masses. The top-quark loop raises $m_h^2$ (Appendix~\ref{app:diagrams}). If the bare parameters must be adjusted to a precision $1/\Delta$ to keep $m_h$ at its observed value, the tuning measure $\Delta\equiv|\delta m_h^2|/m_h^2$ is, for the full combination in Eq.~\eqref{eq:veltman},
\begin{equation}
\Delta\simeq\left(\frac{\Lambda}{552\ {\rm GeV}}\right)^2
\label{eq:tuning}
\end{equation}
for the central values of the Particle Data Group (PDG)~\cite{ParticleDataGroup:2024cfk}. As emphasized recently in Ref.~\cite{Peskin:2025}, the issue is the sensitivity of $m_h$ to any heavy state that couples to the Higgs, rather than the regulator-dependent $\Lambda^2$ itself. In dimensional regularization, for instance, a state of mass $M_X$ contributes $\delta m_h^2\propto M_X^2$. A framework that addresses the problem must therefore control the contributions of heavy states, not only remove a power divergence.

Proposed solutions include supersymmetry~\cite{Draper:2016pys}, composite and pseudo-Nambu--Goldstone Higgs models~\cite{Kaplan:1983fs,Agashe:2004rs,Panico:2015jxa}, neutral naturalness~\cite{Chacko:2005pe}, cosmological selection or relaxation of the weak scale~\cite{Graham:2015cka,Arkani-Hamed:2016rle,Arkani-Hamed:2020yna,TitoDAgnolo:2021nhd,Giudice:2021viw}, ultraviolet/infrared (UV/IR) mixing~\cite{Craig:2019zbn}, classical scale invariance~\cite{Meissner:2006zh,Trautner:2025} and asymptotic safety~\cite{Shaposhnikov:2009pv}. A distinct line of work softens loop integrals with momentum-dependent form factors, such as the Lee--Wick SM~\cite{Lee:1969fy,Lee:1970iw,Grinstein:2007mp}, nonlocal (infinite-derivative) extensions of the SM~\cite{Biswas:2014yia}, ``asymptotic nonlocality''~\cite{Boos:2021chb,Boos:2021jih,Boos:2021lsj}, the quantum-critical Higgs~\cite{Bellazzini:2015cgj,Shirazi:2019xjd} and Higgsplosion~\cite{Khoze:2017tjt}. For Higgsplosion, Refs.~\cite{Belyaev:2018mtd,Monin:2018cbi} showed that an arbitrarily fast suppression of the Higgs propagator conflicts with unitarity and with the spectral representation of local quantum field theory (QFT), a conclusion disputed in Ref.~\cite{Khoze:2018qhz} on the basis of strictly localizable QFT.

In hadron physics, self-similar structures, hadrons made of hadrons, underlie the statistical bootstrap~\cite{Hagedorn:1965st,Frautschi:1971ij,Chew:1970wp}. Replacing Boltzmann--Gibbs by Tsallis statistics~\cite{Tsallis:1987eu,Curado:1991jc}, a one-parameter generalization characterized by an entropic index $q$, describes transverse-momentum spectra in high-energy collisions~\cite{Bediaga:1999hv,Cleymans:2011in,Marques:2012spa}. Tsallis distributions follow from fluctuations of the inverse temperature, an approach known as superstatistics~\cite{Wilk:1999dr,Beck:2003}, and from self-similar ``thermofractal'' structures~\cite{Deppman:2016fxs,Deppman:2019yzp}. A thermofractal $q$-exponential coupling has also been used to regularize the Nambu--Jona-Lasinio (NJL) gap equation~\cite{Megias:2022}.

In this paper we consider a Higgs boson with a self-similar internal structure and examine to what extent this structure can soften the quantum corrections to its mass. When $1/(q-1)$ is an integer, the $q$-exponential form factor $F_q$ derived below coincides with the equal-mass case of asymptotic nonlocality~\cite{Boos:2021chb}, and the present work differs from it by placing the form factor only in the interactions, which preserves the positivity of the Higgs propagator at leading order, and by studying how the tuning and the decoupling of heavy states depend on $q$. Section~\ref{sec:spectral} uses the spectral representation~\cite{Weinberg:1995mt,Peskin:1995ev} to show where internal structure can enter. Section~\ref{sec:formfactor} derives the $q$-exponential form factor from a superstatistical model of the self-similar structure, and Sec.~\ref{sec:lagrangian} gives the effective Lagrangian. Section~\ref{sec:mass} computes the one-loop corrections to the Higgs mass, Sec.~\ref{sec:pheno} discusses naturalness and observables, and Sec.~\ref{sec:rg} the renormalization-group (RG) running. Section~\ref{sec:uv} comments on a confining UV completion and Sec.~\ref{sec:conclusions} concludes. The technical details are collected in the appendices, which cover the spectral representation and positivity (Appendix~\ref{app:spectral}), superstatistics (Appendix~\ref{app:superstat}), the Feynman rules and the diagram-by-diagram self-energy (Appendix~\ref{app:diagrams}), the Higgs mass term and self-couplings (Appendix~\ref{app:trilinear}), the master integral (Appendix~\ref{app:master}), the off-shell estimate (Appendix~\ref{app:offshell}) and the beta-function suppression factor (Appendix~\ref{app:beta}).

%=====================================================================
\section{Spectral representation and internal structure}\label{sec:spectral}
%=====================================================================
For any interacting scalar field $\phi$ the K\"all\'en--Lehmann representation~\cite{Peskin:1995ev,Weinberg:1995mt} expresses the two-point function as a positive superposition of free propagators (Appendix~\ref{app:spectral}). In momentum space,\footnote{We use natural units, $\hbar=c=1$, and the metric $(+,-,-,-)$.}
\begin{align}
\int\!\dd^4x\,e^{ip\cdot x}&\langle\Omega|T\phi(x)\phi(0)|\Omega\rangle=\frac{i\mathcal{Z}}{p^2-m_h^2+i\epsilon}\nonumber\\
&+\int_{s_0}^\infty\!\frac{\dd\sigma}{2\pi}\frac{i\rho(\sigma)}{p^2-\sigma+i\epsilon},
\label{eq:pole}
\end{align}
where $|\Omega\rangle$ is the vacuum, $\rho(\sigma)\geq0$ the spectral density, $s_0$ the lowest multiparticle threshold and $\mathcal{Z}=|\langle\Omega|\phi(0)|h_{\bf 0}\rangle|^2$ the field-strength renormalization, with $|h_{\bf 0}\rangle$ the one-Higgs state at rest. Two consequences are central to this paper.

First, \emph{the residue $\mathcal{Z}$ is a number}. Lorentz invariance gives $\langle\Omega|\phi(0)|h_{\bf p}\rangle=\langle\Omega|\phi(0)|h_{\bf 0}\rangle$ for every exact one-particle state, and hence $\mathcal{Z}$ cannot depend on an external or loop momentum. A Higgs made of many internal configurations, each with its own overlap $\mathcal{Z}_i$ with the field, has $\mathcal{Z}$ equal to a weighted average of the $\mathcal{Z}_i$ for a statistical ensemble of configurations, and to a definite number including interference terms for a single pure state. In neither case does momentum dependence arise (Appendix~\ref{app:spectral}).

Second, \emph{positivity forbids a softer propagator}. Because $\rho\geq0$, the two-point function cannot vanish faster than $1/p^2$ as $|p^2|\to\infty$~\cite{Weinberg:1995mt}. A propagator falling as $1/p^4$ necessarily contains a negative-norm state (a ghost), as shown explicitly in Appendix~\ref{app:spectral}. For a canonically normalized field $0\leq \mathcal{Z}\leq1$, and $\mathcal{Z}=0$ characterizes a composite particle~\cite{Weinberg:1995mt,Weinberg:1965zz}.

Internal structure that respects positivity can therefore soften loop integrals only through the \emph{vertex functions} that couple the composite Higgs to other fields, exactly as the electromagnetic form factor of the proton softens its self-energy. In the effective Lagrangian of Sec.~\ref{sec:lagrangian} this keeps the Higgs propagator of the form~\eqref{eq:pole} at leading order. After electroweak symmetry breaking the interactions of Sec.~\ref{sec:lagrangian} induce corrections of relative order $m_h^2/M^2$, where $M$ is the compositeness scale introduced in Sec.~\ref{sec:formfactor}. These corrections are discussed in Sec.~\ref{sec:lagrangian}. We describe the coupling of the Higgs (momentum $p$) to a pair of fields $\Phi_a$, $\Phi_b$ with momenta $p/2\pm k$ by an amputated vertex
\begin{equation}
G_{ab}(k;p)=g_{ab}\,F(k_E^2),\qquad F(0)=1,
\label{eq:vertex}
\end{equation}
where $g_{ab}$ is the corresponding SM coupling, $k$ is the relative momentum of the pair and $k_E$ is its Euclidean counterpart, with $k_E^2\equiv-k^2>0$ for spacelike $k$. The normalization $F(0)=1$ defines $g_{ab}$ as the coupling at zero relative momentum, where a long-wavelength probe sees a pointlike Higgs and $g_{ab}$ equals the SM coupling, just as the electric form factor of the proton is normalized to its charge. For a general composite the vertex depends on $k^2$, $p^2$ and $k\cdot p$. In the loops relevant here the loop momentum dominates, and $p^2=m_h^2$ gives corrections of relative order $m_h^2/M^2$, which are neglected. In the model of Sec.~\ref{sec:lagrangian} the vertex depends on $k^2$ alone by construction. A one-loop self-energy diagram with one such vertex is then weighted by $F$ and one with two vertices by $F^2$, evaluated at the relative momentum of the pair attached to each vertex (Table~\ref{tab:weights}). The full propagator keeps the form $i\mathcal{Z}/(p^2-m_h^2)$ near the pole, with $m_h^2=m_0^2+\Sigma(m_h^2)$ and $\mathcal{Z}^{-1}=1-\Sigma'(m_h^2)$, where $m_0$ is the bare mass and $\Sigma$ the self-energy.
\begin{table}[t]
\caption{Form-factor weight $w_j$ of each one-loop diagram class $j$ of Fig.~\ref{fig:diagrams}, following from Eq.~\eqref{eq:Lvertex} (with $k$ the loop momentum and $p$ the external momentum), and the resulting effective cutoff $\Lambda_{{\rm eff},j}^2=\int_0^\infty\dd u\,w_j(u)$ of class $j$, $u=k_E^2$. For the loops with one form factor (scalar seagull within one bilinear, gauge seagull) $\Lambda_{{\rm eff},j}=\Lambda_{\rm eff}$ of Eq.~\eqref{eq:Leff}, and for those with two form factors $\Lambda_{{\rm eff},j}^2=M^2/(3-q)$ (top-quark loop) or $4M^2/(3-q)$ (cross-contracted seagull, whose single vertex contains two bilinears). The coefficient $6\lambda$ of the scalar seagull splits as $6\lambda=4\lambda$ within one bilinear $H^\dagger\star_qH$ ($\lambda$ from $h$ and $3\lambda$ from the three Goldstone bosons) $+\,2\lambda$ across the two bilinears ($h$ only, since the background field points along $h$), Eq.~\eqref{eq:dmu} and Appendix~\ref{app:diagrams}.}
\label{tab:weights}
\renewcommand{\arraystretch}{1.3}
\begin{ruledtabular}
\begin{tabular}{lcc}
diagram & weight $w_j$ & $\Lambda_{{\rm eff},j}^2$\\
\hline
(a) seagull, within & $F_q(k_E^2)$ & $M^2/(2-q)$\\
(a) seagull, across & $F_q\big((k+p)_E^2/4\big)^2$ & $4M^2/(3-q)$\\
(c) gauge seagull & $F_q(k_E^2)$ & $M^2/(2-q)$\\
(g) top loop & $F_q\big((k-p/2)_E^2\big)^2$ & $M^2/(3-q)$\\
(b), (d), (e) bubbles & $F_q^2$ & (log. in SM)\\
\end{tabular}
\end{ruledtabular}
\end{table}

%=====================================================================
\section{Self-similar structure and the $q$-exponential form factor}\label{sec:formfactor}
%=====================================================================
The coupling of the Higgs to a pair of fields, Eq.~\eqref{eq:vertex}, depends on the relative momentum $k$ of the pair. The relative momentum $k$ is the Fourier conjugate of the separation of the two fields. The vertex form factor $F(k_E^2)$ is therefore the momentum-space counterpart of how the constituents are spread out inside the Higgs, in the same way that the momentum-space wave function of a bound state, or the electromagnetic form factor of the proton, reflects its spatial structure.

Consider first a single configuration of the constituents, spread over a region of squared size $\beta$. If the spatial profile is Gaussian with squared width $\beta$, its Fourier transform is again Gaussian,
$F(k_E^2|\beta)=e^{-\beta k_E^2}$, the simplest localized profile (as for the ground state of a harmonic oscillator). In a single configuration $\beta$ is fixed. This factor has the form of a Boltzmann factor $e^{-E/T}$, with $k_E^2$ in the role of an energy and $\beta$ in that of an inverse temperature. Used alone, it gives the Gaussian form factor of nonlocal field theories~\cite{Biswas:2014yia}.

In a self-similar system the constituents are themselves composite, and the size changes from one configuration to the next and from one level of constituents to the next. Consequently, $\beta$ is distributed over an ensemble of configurations rather than fixed. If $\beta$ is the sum of the squares of the $3N$ Cartesian coordinates of $N$ constituents, each Gaussian distributed, it follows a $\chi^2$ distribution with $3N$ degrees of freedom, i.e.\ the Gamma distribution
\begin{equation}
f(\beta)=\frac{1}{\Gamma(a)}\left(\frac{a}{\langle\beta\rangle}\right)^{a}\beta^{a-1}e^{-a\beta/\langle\beta\rangle},\qquad a=\frac{3N}{2},
\label{eq:gamma}
\end{equation}
where $\Gamma(a)$ is the Euler Gamma function. The distribution has mean $\langle\beta\rangle$ and relative variance $1/a$.\footnote{The identification $a=3N/2$ counts the three rest-frame spatial coordinates of each constituent. Counting four Euclidean coordinates, or removing the centre-of-mass motion, would change the map between $q$ and $N$ (to $a=2N$ or $a=3(N-1)/2$, respectively). $N$ is therefore an effective number of constituents, while the relation $q-1=1/a={\rm var}(\beta)/\langle\beta\rangle^2$ does not depend on this choice.}

Averaging a Boltzmann factor over a fluctuating inverse temperature is known as superstatistics~\cite{Wilk:1999dr,Beck:2003}. It turns the exponential into a power law, the Tsallis $q$-exponential. Here we average the Gaussian form factor over the fluctuating size $\beta$, which gives, exactly (Appendix~\ref{app:superstat}),
\begin{equation}
F_q(k_E^2)=\int_0^\infty\!\dd\beta\,f(\beta)\,e^{-\beta k_E^2}=\left[1+(q-1)\frac{k_E^2}{M^2}\right]^{-\frac{1}{q-1}},
\label{eq:Fq}
\end{equation}
with
\begin{equation}
q-1=\frac{1}{a}=\frac{2}{3N},\qquad M^2\equiv\frac{1}{\langle\beta\rangle}.
\label{eq:qN}
\end{equation}
Thus $q-1$ is the relative variance of $\beta$, and $1/M$ is the average size of the Higgs. We refer to $M$ as the compositeness scale. For every $q$, the derivative of $F_q$ with respect to $k_E^2$ at $k_E^2=0$ equals $-\langle\beta\rangle=-1/M^2$. The compositeness scale $M$ is therefore fixed by the mean squared size $\langle\beta\rangle$, just as a charge radius is defined from the derivative of a form factor at zero momentum. Different $q$ at the same $M$ thus share the same mean size and differ only in its fluctuations. Equation~\eqref{eq:Fq} is the Tsallis $q$-exponential, $F_q(k_E^2)=e_q(-k_E^2/M^2)$ with $e_q(x)\equiv[1+(1-q)x]^{1/(1-q)}$. The relation $q-1=2/(3N)$ coincides with the member $\nu=0$ of the thermofractal family $q-1=2(1-\nu)/(3N)$ of Ref.~\cite{Deppman:2016fxs}, where $\nu$ is the power with which the energy distribution of a subsystem feeds back on itself.

\begin{figure}[t]
\centering
\includegraphics[width=0.95\columnwidth]{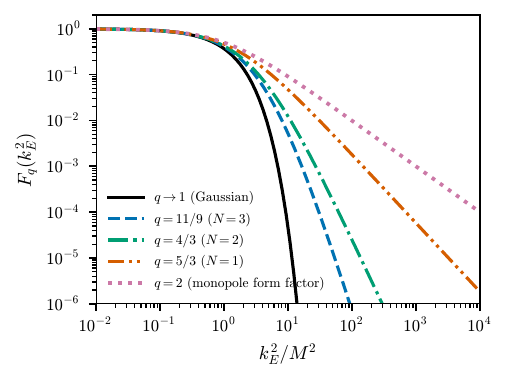}
\caption{The self-similar form factor $F_q(k_E^2)$ of Eq.~\eqref{eq:Fq} for the benchmarks $q=11/9$, $4/3$, $5/3$ ($N=3,2,1$), the Gaussian limit $q\to1$ and the monopole form factor $q=2$.}
\label{fig:ff}
\end{figure}

Integer numbers of constituents give the benchmarks used below, $q=5/3$, $4/3$ and $11/9$ for $N=1$, $2$ and $3$,\footnote{$N$ counts the constituents at each level of the self-similar structure, each constituent again having $N$ constituents, much as a fractal is built from a finite number of pieces at each step. Including the recursive feedback of the thermofractal, $\nu\neq0$ in Ref.~\cite{Deppman:2016fxs}, keeps the $q$-exponential form and only changes the map between $q$ and $N$.} while $q\to1$ corresponds to $N\to\infty$.
\begin{itemize}
\item For $q\to1$, $F_q\to e^{-k_E^2/M^2}$, the Gaussian form factor of the nonlocal SM~\cite{Biswas:2014yia}. For $q=2$ it reduces to a monopole form factor, $F_2=M^2/(M^2+k_E^2)$. With the exponent $n\equiv1/(q-1)$ of Eq.~\eqref{eq:Fq}, $q=3/2$ gives $n=2$ and $F_{3/2}=[1+k_E^2/(2M^2)]^{-2}$ is a dipole form factor, the shape that approximately describes the electromagnetic form factors of the proton~\cite{Perdrisat:2006hj}.
\item The region $k_E^2\gg M^2$ probes distances much shorter than the average size $1/M$, i.e.\ it resolves the constituents, as in deep inelastic scattering, where a large momentum transfer resolves the partons inside the proton~\cite{Peskin:1995ev}. For a single Gaussian configuration the form factor would be exponentially small there. After averaging over sizes it falls only as a power law, $F_q\simeq[M^2/((q-1)k_E^2)]^{n}$, because small configurations dominate at large momentum. This power-law tail reflects the self-similar structure. In a loop, $k_E^2\gg M^2$ is the ultraviolet region, and each insertion lowers the superficial degree of divergence by $2n$.
\item $F_q^2$ is again a $q$-exponential form factor, $F_q(k_E^2)^2=F_{\tilde q}(k_E^2)|_{M^2\to M^2/2}$ with $\tilde q=(q+1)/2$, so that loops with two vertices follow from the formulas for one vertex.
\item Although loop integrals are evaluated in Euclidean space, physical processes also involve timelike relative momenta (decays, $s$-channel production), which require the analytic continuation $k_E^2\to-k^2$. In Minkowski space, $F_q(k^2)=[1-(q-1)k^2/M^2]^{-n}$ grows for timelike $k^2>0$ and is singular at $k^2=M^2/(q-1)$. The singularity is a simple pole for $q=2$ (a Lee--Wick-type partner~\cite{Lee:1969fy,Lee:1970iw,Grinstein:2007mp}), a pole of order $n$ for integer $n$ (the equal-mass case of asymptotic nonlocality~\cite{Boos:2021chb}), and a branch point otherwise. These timelike singularities signal the new states of the composite sector. The effective description breaks down as they are approached.
\end{itemize}
Figure~\ref{fig:ff} compares the members of the family.

%=====================================================================
\section{Effective Lagrangian}\label{sec:lagrangian}
%=====================================================================
The interactions of a composite Higgs with a form factor are necessarily nonlocal. Each term of a local Lagrangian, with finitely many derivatives, gives a vertex that is a polynomial in the momenta, whereas $F_q$ is not a polynomial. In position space a form factor means that the fields entering a vertex interact at separations of order $1/M$, the size of the composite, just as the electromagnetic coupling of the proton is spread over its charge radius although QCD is local. The framework is therefore a nonlocal effective field theory with an emergent nonlocality scale $1/M$, not a fundamentally nonlocal theory. The underlying confining theory (Sec.~\ref{sec:uv}) is local. Following Sec.~\ref{sec:spectral}, we place the form factor in the interactions. The kinetic terms are local, and the interactions of the Higgs doublet $H$ are smeared over the size of the composite, i.e.\ nonlocal. The Higgs propagator therefore keeps its positive spectral representation at leading order. Define the nonlocal product
\begin{equation}
(A\star_q B)(x)=\int\!\dd^4y\,\mathcal{K}_q(y)\,A\!\left(x+\tfrac{y}{2}\right)\mathcal{U}\!\left(x+\tfrac{y}{2},x-\tfrac{y}{2}\right)B\!\left(x-\tfrac{y}{2}\right),
\label{eq:star}
\end{equation}
where the kernel $\mathcal K_q(y)$ is the four-dimensional Fourier transform of $F_q$ and $\mathcal U$ is a Wilson line that keeps the product gauge covariant~\cite{Terning:1991yt}. If $A$ and $B$ carry incoming momenta $p_A$ and $p_B$, the $y$ integral produces $F_q(k_E^2)$ with $k=(p_A-p_B)/2$, the relative momentum of the pair (Appendix~\ref{app:diagrams}). The effective Lagrangian is
\begin{align}
\mathcal L=&-\tfrac14 F^a_{\mu\nu}F^{a\mu\nu}+\sum_f\bar f\, i\slashed{D}\,f+(D_\mu H)^\dagger D^\mu H\nonumber\\
&-m_H^2H^\dagger H-\lambda\,(H^\dagger\star_q H)(H^\dagger\star_q H)\nonumber\\
&-y_t\!\int\!\dd^4y\,\mathcal K_q(y)\,\bar Q_L\!\left(x+\tfrac y2\right)\tilde H(x)\,t_R\!\left(x-\tfrac y2\right)\nonumber\\
&+{\rm h.c.}+\mathcal L_{HV}^{(q)},
\label{eq:Lvertex}
\end{align}
where $F^a_{\mu\nu}$ are the field strengths of the gauge fields, with $a$ running over the generators of SU(3)$\times$SU(2)$\times$U(1), and $f$ runs over the SM fermions. The covariant derivative of the Higgs doublet is $D_\mu=\partial_\mu-ig\,W^a_\mu T^a-ig'\,YB_\mu$, where $W^a_\mu$ ($a=1,2,3$) and $B_\mu$ are the SU(2) and U(1) gauge fields, $g$ and $g'$ their couplings, $T^a=\sigma^a/2$ the SU(2) generators, with $\sigma^a$ the Pauli matrices, and $Y=1/2$ the hypercharge of $H$. Further, $Q_L$ and $t_R$ are the third-generation quark doublet and right-handed top, $y_t=\sqrt2\,m_t/v$ is the top Yukawa coupling,\footnote{Since the $\star_q$ product is defined for two fields, the Yukawa term is written explicitly, with the quark pair at $x\pm y/2$ carrying $F_q$ and the Higgs at the centre point $x$.} $\tilde H=i\sigma^2H^*$, and the Wilson lines in the Yukawa term are implicit. In the local limit $F_q=1$ the Higgs potential is $V(H)=m_H^2H^\dagger H+\lambda(H^\dagger H)^2$, with $m_H^2=-\lambda v^2<0$ and $m_h^2=2\lambda v^2=-2m_H^2$ at tree level. We denote by $h$ the physical Higgs field and by $G^{0,\pm}$ the would-be Goldstone bosons. Other Yukawa couplings are treated in the same way and are negligible here. The Higgs kinetic term is local. As a working definition, $\mathcal L^{(q)}_{HV}$ replaces, in the terms of $(D_\mu H)^\dagger D^\mu H$ quadratic in the gauge fields, the products $W^a_\mu W^{a\mu}$, $B_\mu B^\mu$ and $W^a_\mu B^\mu$ by their $\star_q$ products, so that the $HVV$ and $HHVV$ vertices ($V=W,Z,\gamma$) carry $F_q$ of the relative momentum of the two gauge bosons. This definition is in tension with the gauge invariance of a local $|D_\mu H|^2$, although the gauge-seagull contribution that relies on it is smaller than the top-quark loop for all benchmarks (Sec.~\ref{sec:mass}). Constructing the smeared gauge interactions gauge invariantly, with Wilson lines as in Ref.~\cite{Terning:1991yt}, is the main open problem of the effective Lagrangian. That such constructions can regulate gauge loops consistently is shown by the gauge-invariant nonlocal models of Ref.~\cite{Boos:2021jih}.

After electroweak symmetry breaking the Higgs propagator receives a small correction, which arises in three steps. (i) With $H\to(0,(v+h)/\sqrt2)^T$, the quartic term also produces a term quadratic in $h$, which contributes to the Higgs mass. Its form factors depend on the Higgs momentum $p$, and the mass therefore becomes momentum dependent (Appendix~\ref{app:trilinear}),
\begin{equation}
m^2(p^2)=\lambda v^2\left[F_q(p^2)+2F_q\!\left(\tfrac{p^2}{4}\right)^2-1\right]\simeq m_h^2\left(1+\frac{p^2}{M^2}\right),
\label{eq:m2main}
\end{equation}
with $F_q$ in its Minkowski form. (ii) The tree-level propagator $i/[p^2-m^2(p^2)]$ then has residue $\mathcal{Z}=[1-\dd m^2/\dd p^2]^{-1}_{p^2=m_h^2}\simeq1+m_h^2/M^2>1$, where $\lambda$ is retuned to $\lambda_{\rm SM}(1-m_h^2/M^2)$, with $\lambda_{\rm SM}=m_h^2/(2v^2)$, to keep the pole at $m_h$. (iii) Positivity requires $0\leq \mathcal{Z}\leq1$ (Sec.~\ref{sec:spectral}), since the propagator still falls as $1/p^2$ (Appendix~\ref{app:trilinear}). Consequently, a negative spectral weight of order $m_h^2/M^2$ appears near the timelike singularities of $F_q$, the nearest at $p^2=M^2/(q-1)$ (Appendix~\ref{app:trilinear}). Numerically, $\mathcal{Z}-1$ is $1.6\%$ at $M=1$~TeV and decreases as $1/M^2$. This differs in kind from the softened propagator excluded in Sec.~\ref{sec:spectral}. A propagator falling as $1/p^4$ violates positivity at leading order, whereas here the violation is an $\mathcal O(m_h^2/M^2)$ effect at the scale of the new states. In hadron physics the timelike singularities of form factors are physical resonances with positive spectral weight. The pion electromagnetic form factor, for instance, is dominated by the $\rho$-meson pole (vector-meson dominance)~\cite{OConnell:1995wf}. The negative weight found here signals that the effective form factor does not yet describe the new states at $k^2=M^2/(q-1)$ unitarily. A complete theory must replace these singularities by states with positive weight. We regard this as part of the open problem of completing the effective Lagrangian, together with gauge invariance. The kernel $\mathcal K_q$ is positive in Euclidean space, since it is a superposition of Gaussians, and the theory is defined there and continued analytically to Minkowski space. The timelike singularities of $F_q$ then raise the usual questions of unitarity and causality of nonlocal and Lee--Wick theories~\cite{Pius:2016jsl,Briscese:2018oyx,Anselmi:2018kgz,Donoghue:2019fcb}.

The form factor could instead be placed in the propagators, through a nonlocal kinetic operator, which gives the nonlocal SM~\cite{Biswas:2014yia} for $q\to1$ and the Lee--Wick-type theories of Sec.~\ref{sec:formfactor}. This violates positivity at leading order for every field (Appendix~\ref{app:spectral}), and we therefore do not adopt it.

%=====================================================================
\section{One-loop corrections to the Higgs mass}\label{sec:mass}
%=====================================================================
\begin{figure*}[t]
\centering
\includegraphics[width=0.95\textwidth]{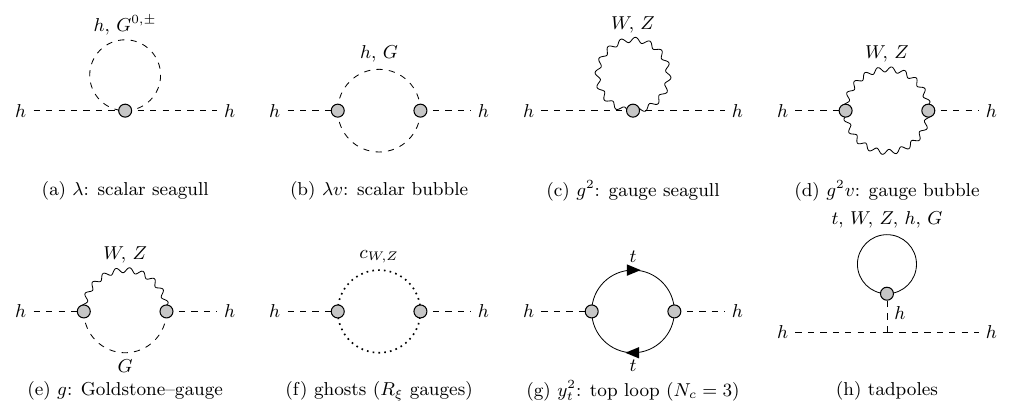}
\caption{One-loop contributions to the Higgs self-energy. Shaded blobs are vertices carrying the self-similar form factor $F_q$. In the SM, diagrams (a), (c) and (g) carry the quadratic divergence of Eq.~\eqref{eq:veltman}, while (b), (d), (e) and (f) are at most logarithmically divergent. The ghost loops (f) vanish in Landau gauge, and the tadpoles (h) shift the vacuum and fix the sign of the physical mass shift.}
\label{fig:diagrams}
\end{figure*}

The one-loop correction to the Higgs mass follows from the one-loop effective potential~\cite{Coleman:1973jx}, the vacuum energy of the fields in a constant Higgs background $\phi$,
\begin{equation}
V_{\rm 1\,loop}(\phi)=\frac12\sum_jn_j\int\!\frac{\dd^4k_E}{(2\pi)^4}\,w_j(k_E^2)\ln\!\big[k_E^2+m_j^2(\phi)\big].
\label{eq:Veff}
\end{equation}
Equation~\eqref{eq:Veff} holds at leading order. With nonlocal interactions the field-dependent masses inside the logarithm also depend on momentum, which affects only the finite terms $-m^2J_q$ below. The sum runs over the contraction classes $j$ of Table~\ref{tab:weights}, the $h$ loop within one bilinear and across the two bilinears, the Goldstone loops $G$ and the $W$, $Z$ and $t$ loops, with field-dependent masses $m_j^2(\phi)$ and degrees of freedom $n_h=1$ for each of the two $h$ classes, $n_G=3$, $n_W=6$ and $n_Z=3$ (three polarizations in Landau gauge, in which the Goldstone bosons are massless at the minimum and the ghosts decouple from $h$), and $n_t=-4N_c=-12$ (negative for fermions, with $N_c=3$ colours). For the two $h$ classes $m_j^2(\phi)$ is the part of the $h$ mass, $3\lambda\phi^2-\lambda v^2$, that arises within ($\partial m_j^2/\partial\phi=2\lambda\phi$) or across ($4\lambda\phi$) the bilinears (Appendix~\ref{app:diagrams}). The weight $w_j$ is the form factor carried by the loops of class $j$ (Table~\ref{tab:weights}). It is one form factor, $F_q$, for the seagull loops within one bilinear and the gauge seagull, and two form factors for the top-quark loop and the cross-contracted seagull, whereas $w_j=1$ in the SM. The loops shift the coefficient of $H^\dagger H$ by $\delta m_H^2=v^{-1}\partial V_{\rm 1\,loop}/\partial\phi|_{\phi=v}$, and the physical mass by $\delta m_h^2=-2\,\delta m_H^2$, since $m_h^2=-2m_H^2$ at tree level. Because $\partial\ln(k_E^2+m^2)/\partial m^2=1/(k_E^2+m^2)$,
\begin{equation}
\delta m_h^2=-\frac{1}{16\pi^2v}\sum_jn_j\frac{\partial m_j^2}{\partial\phi}\bigg|_v I_j,
\label{eq:CW}
\end{equation}
where $I_j$ is the master integral below with the weight $w_j$. For a loop particle of mass $m$ and the weight $F_q$ it reads
\begin{align}
I_q(m^2;M^2)&\equiv16\pi^2\!\int\!\frac{\dd^4k_E}{(2\pi)^4}\frac{F_q(k_E^2)}{k_E^2+m^2}\nonumber\\
&=\int_0^\infty\!\dd u\,\frac{u\,F_q(u)}{u+m^2}\nonumber\\
&=\Lambda_{\rm eff}^2-m^2J_q(m^2;M^2),
\label{eq:master}
\end{align}
with $u\equiv k_E^2$. Writing $u/(u+m^2)=1-m^2/(u+m^2)$ splits $I_q$ into $\Lambda_{\rm eff}^2-m^2J_q$, where the first term is the piece that is quadratically divergent in the SM (with $w=1$ and a hard cutoff it equals $\Lambda^2$), and $\Lambda_{\rm eff}^2$ therefore plays the role of $\Lambda^2$ in Eq.~\eqref{eq:veltman}, which allows a direct comparison with Eq.~\eqref{eq:tuning}. For $1<q<2$ (Appendix~\ref{app:master}) one finds
\begin{align}
\Lambda_{\rm eff}^2&=\int_0^\infty\!\dd u\,F_q(u)=\frac{M^2}{2-q},\label{eq:Leff}\\
J_q(m^2;M^2)&=\int_0^\infty\!\frac{\dd u\,F_q(u)}{u+m^2}\nonumber\\
&\simeq\ln\frac{M^2}{(q-1)m^2}-\psi\!\left(n\right)-\gamma_E .
\label{eq:J}
\end{align}
Here $\psi(a)=\Gamma'(a)/\Gamma(a)$ is the digamma function and $\gamma_E\simeq0.5772$ is Euler's constant. The last form holds for $m^2\ll M^2$. Appendix~\ref{app:master} gives the exact expression in terms of a hypergeometric function. For $q\to1$ one recovers $\Lambda_{\rm eff}^2=M^2$ and $J=\ln(M^2/m^2)-\gamma_E$, i.e.\ the nonlocal result $\delta m^2=\lambda M^2/(32\pi^2)$ of Refs.~\cite{Biswas:2014yia,Boos:2021chb}. The superstatistical average is taken independently at each vertex (at each bilinear for the cross-contracted seagull), which gives $F_q^2$ for loops with two form factors. The top-quark loop then follows from the same formulas with $(q,M^2)\to(\tilde q,M^2/2)$, which gives the effective cutoff $M^2/(3-q)$. The cross-contracted seagull follows with $(q,M^2)\to(\tilde q,2M^2)$, which gives $4M^2/(3-q)$. Averaging with one configuration for the whole loop changes the results only by factors of order one (Appendix~\ref{app:superstat}).

Summing over the species with the weights of Table~\ref{tab:weights} (Appendix~\ref{app:diagrams}), Eq.~\eqref{eq:CW} gives
{\interdisplaylinepenalty=10000
\begin{align}
8\pi^2v^2\,\delta m_h^2={}&-\Big[\tfrac12m_h^2\big[I_q(m_h^2;M^2)+3I_q(0;M^2)\big]\nonumber\\
&+m_h^2\,I_{\tilde q}(m_h^2;2M^2)\nonumber\\
&+6m_W^2\,I_q(m_W^2;M^2)\nonumber\\
&+3m_Z^2\,I_q(m_Z^2;M^2)\nonumber\\
&-12m_t^2\,I_{\tilde q}(m_t^2;M^2/2)\Big],
\label{eq:dmu}
\end{align}}%
where the first two lines are the Higgs and Goldstone loops (within and across the bilinears). Keeping only the $M^2$ terms,
\begin{align}
|\delta m_h^2|\simeq\frac{1}{8\pi^2v^2}\bigg|&\left(2m_h^2+6m_W^2+3m_Z^2\right)\frac{M^2}{2-q}\nonumber\\
&+4\left(m_h^2-3m_t^2\right)\frac{M^2}{3-q}\bigg|.
\label{eq:dmh}
\end{align}
Parametrically, Eq.~\eqref{eq:dmh} is Veltman's estimate with the cutoff replaced by the compositeness scale, as for any composite Higgs, and the new elements are the dependence on $q$, the different effective cutoffs of the loops with one and two form factors and their partial cancellation, and the decoupling of heavy states. Compared with Eq.~\eqref{eq:veltman}, the loops with one form factor (scalar seagull within one bilinear, gauge seagull) have the effective cutoff $M^2/(2-q)$, whereas those with two form factors have $M^2/(3-q)$ (top-quark loop) or $4M^2/(3-q)$ (cross-contracted seagull, whose single vertex contains two bilinears). The gauge-seagull contribution, which relies on the working definition of Sec.~\ref{sec:lagrangian}, amounts to $40$--$79\%$ of the top-quark loop at $M=1$~TeV for $q\to1$ to $q=5/3$. The correction is finite for $1<q<2$. For $q=2$ the loops with one form factor diverge logarithmically while those with two remain finite, and for $q>2$ the former diverge as $\Lambda^{2(q-2)/(q-1)}$. No cutoff appears for $1<q<2$, where the form factor itself makes all integrals finite and cuts the loops off at $\Lambda_{{\rm eff},j}^2$ (Table~\ref{tab:weights}). A hard cutoff $\Lambda$ is needed for $q\geq2$, and at $q=2$ the loops follow from Eq.~\eqref{eq:I2}.

\paragraph*{Heavy states.} A heavy complex scalar $X$ of mass $M_X$ coupled through $\lambda_X(H^\dagger\star_qH)(X^\dagger\star_qX)$ contributes $\delta m_h^2=-\lambda_X I_q(M_X^2;M^2)/(8\pi^2)$. For $M_X\gg M$ and $1<q<2$ (Appendix~\ref{app:master}),
\begin{align}
I_q(M_X^2;M^2)&\simeq\frac{M^4}{(2-q)(3-2q)\,M_X^2}\quad(q<\tfrac32),\nonumber\\
I_q(M_X^2;M^2)&\simeq\frac{4M^4}{M_X^2}\ln\frac{M_X^2}{M^2}\quad(q=\tfrac32),\label{eq:heavy}\\
I_q(M_X^2;M^2)&\simeq\frac{\pi\,(q-1)^{-n}}{|\sin(\pi n)|}\,\frac{M^{2n}}{M_X^{2n-2}}\quad(q>\tfrac32),\nonumber
\end{align}
Instead of growing like $M_X^2$, heavy contributions decrease with $M_X$, the criterion of Ref.~\cite{Peskin:2025}, provided $X$ couples to the Higgs through the smeared vertices. For a heavy fermion coupled through two Yukawa vertices, Eq.~\eqref{eq:heavy} applies with $(q,M^2)\to(\tilde q,M^2/2)$. Since $\tilde q<3/2$, its first line always applies. At $q=2$, by contrast, the seagull with a heavy state in the loop behaves as $M^2\ln(\Lambda^2/M_X^2)$. Its $M_X$-dependent part, $I_2(M_X^2;M^2)-I_2(0;M^2)$, shifts $m_h^2$ by $\lambda_X M^2\ln(M_X^2/M^2)/(8\pi^2)$, which does not decouple.

%=====================================================================
\section{Naturalness and observables}\label{sec:pheno}
\begin{figure}[t]
\centering
\includegraphics[width=0.95\columnwidth]{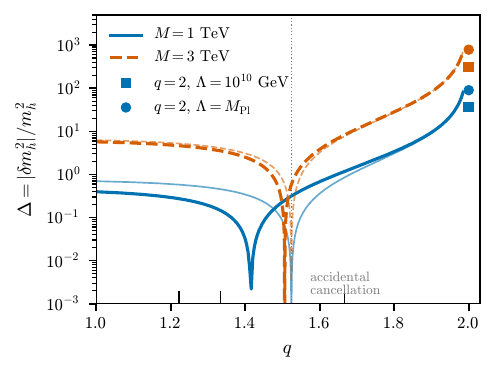}
\caption{Tuning $\Delta=|\delta m_h^2|/m_h^2$ as a function of $q$ for $M=1$ and $3$~TeV, from Eq.~\eqref{eq:dmu} (thick lines) and from the leading $M^2$ terms of Eq.~\eqref{eq:dmh} (thin lines). The dotted line marks the accidental cancellation of the $M^2$ terms. Ticks on the horizontal axis mark the benchmarks. Symbols at $q=2$ show the logarithmically divergent case for two values of the cutoff. The downward spikes are zeros of $\delta m_h^2$ (sign changes from the competition of boson and top loops).}
\label{fig:tuning}
\end{figure}

%=====================================================================
\paragraph*{Tuning.} We measure the tuning by $\Delta=|\delta m_h^2|/m_h^2$, with $\delta m_h^2$ from Eq.~\eqref{eq:dmu}. Equation~\eqref{eq:dmu} contains the leading $M^2$ terms and the finite terms $-m^2J_q$ of each loop of mass $m$. It does not contain all finite one-loop terms, since the dependence on the external momentum and the bubble diagrams are omitted (Appendix~\ref{app:diagrams}). These omitted terms are small compared with the $M^2$ terms when $M$ is large. When $\Delta\lesssim1$, however, the $M^2$ terms are themselves small and the omitted terms can be equally large. The values of $\Delta$ below are therefore estimates of the order of magnitude rather than precise numbers. The omitted terms are not needed for the conclusions, since they matter only when $\Delta\lesssim1$, where no tuning is required whatever their precise value, and the finite parts of the gauge bubbles would require the gauge-invariant completion of Sec.~\ref{sec:lagrangian}. Figure~\ref{fig:tuning} shows $\Delta$ as a function of $q$ for $M=1$ and $3$~TeV. For the benchmarks there is little or no tuning, $\Delta\lesssim1$, for $M$ up to about $1$--$2$~TeV, depending on $q$ ($\Delta=0.27$, $0.15$ and $1.2$ at $M=1$~TeV for $q=11/9$, $4/3$ and $5/3$), and in this range $\Delta$ is set by the finite terms. The largest $M$ compatible with a given $\Delta$ is listed in Table~\ref{tab:Mmax}. In particular, $\Delta\leq10$ allows $M$ up to $3.3$--$4.8$~TeV. For $q=2$ the seagulls diverge logarithmically, $\delta m_h^2\propto M^2\ln(1+\Lambda^2/M^2)$, Eq.~\eqref{eq:I2}, and require a hard cutoff $\Lambda$. Table~\ref{tab:q2} lists $\Delta$ for $q=2$ with $\Lambda=10$~TeV, $10^{10}$~GeV and the Planck mass $M_{\rm Pl}\simeq1.22\times10^{19}$~GeV, together with the cutoff-independent benchmarks, and the symbols in Fig.~\ref{fig:tuning} show the $q=2$ values for $\Lambda=10^{10}$~GeV and $M_{\rm Pl}$. With $\Lambda=M_{\rm Pl}$ one finds $\Delta\simeq33$--$58$ for $M=0.6$--$0.8$~TeV, compared with $\Delta\lesssim1$ for the benchmarks at the same $M$. A logarithmic divergence alone therefore does not make the Higgs mass natural.

The leading $M^2$ term of Eq.~\eqref{eq:dmh} changes sign at $q\simeq1.52$, where the boson loops, which grow as $1/(2-q)$, cancel the top loop, which grows as $1/(3-q)$. This is an accidental Veltman-type cancellation, not a prediction of the framework. Close to this value of $q$ the tuning $\Delta$ becomes very small, and its actual value is then set by the finite terms that Eq.~\eqref{eq:dmu} does not contain. The finite terms included in Eq.~\eqref{eq:dmu} shift the zero to $q\simeq1.41$ at $M=1$~TeV, and at $M\simeq0.73$~TeV the zero falls on the benchmark $q=4/3$. The very small value $\Delta\approx0.01$ in Table~\ref{tab:q2} is therefore a coincidence, not a prediction. We keep $q=4/3$ as a benchmark because it corresponds to $N=2$, independently of this cancellation.

\begin{table}[t]
\caption{Largest compositeness scale $M_{\max}$ (TeV) compatible with a tuning $\Delta$, i.e.\ the largest $M$ for which the tuning does not exceed the $\Delta$ of that column, from Eq.~\eqref{eq:dmu}, and the corresponding on-shell coupling shift $1-\kappa$ at $\Delta=10$, where $\kappa\simeq1-m_h^2/(4M^2)$ is the on-shell coupling modifier of Sec.~\ref{sec:pheno}.}
\label{tab:Mmax}
\begin{ruledtabular}
\begin{tabular}{lcccc}
$q$ & $\Delta=1$ & $\Delta=10$ & $\Delta=100$ & $1-\kappa$ ($\Delta=10$)\\
\hline
$\to1$ & 1.41 & 3.91 & 12.0 & $2.6\times10^{-4}$\\
$11/9$ & 1.56 & 4.29 & 13.2 & $2.1\times10^{-4}$\\
$4/3$ & 1.78 & 4.84 & 14.9 & $1.7\times10^{-4}$\\
$5/3$ & 0.91 & 3.31 & 10.8 & $3.6\times10^{-4}$\\
\end{tabular}
\end{ruledtabular}
\end{table}

\begin{table}[t]
\caption{Tuning $\Delta$ for $q=2$ with a hard cutoff $\Lambda$ and for the benchmarks $1<q<2$ (cutoff independent), as a function of $M$.}
\label{tab:q2}
\begin{ruledtabular}
\begin{tabular}{llcccc}
 & & \multicolumn{4}{c}{$M$ (TeV)}\\
\cline{3-6}
$q$ & $\Lambda$ & $0.5$ & $0.75$ & $1$ & $3$\\
\hline
$2$ & $10$~TeV & $1.1$ & $1.8$ & $2.3$ & $6.3$\\
$2$ & $10^{10}$~GeV & $9.9$ & $21$ & $37$ & $310$\\
$2$ & $M_{\rm Pl}$ & $23$ & $51$ & $90$ & $785$\\
$11/9$ & --- & $0.02$ & $0.08$ & $0.27$ & $4.7$\\
$4/3$ & --- & $0.06$ & $0.01$ & $0.15$ & $3.5$\\
$5/3$ & --- & $0.39$ & $0.74$ & $1.2$ & $8.3$\\
\end{tabular}
\end{ruledtabular}
\end{table}

\begin{figure}[t]
\centering
\includegraphics[width=0.95\columnwidth]{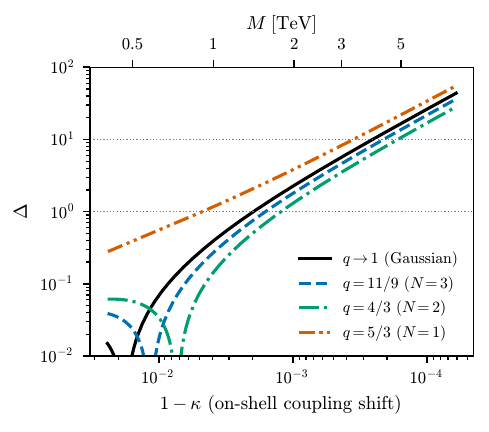}
\caption{Tuning $\Delta$ versus the on-shell coupling shift $1-\kappa\simeq m_h^2/(4M^2)$ for the benchmarks. The horizontal axis is reversed so that $M$, shown on the upper axis, increases to the right. The dips are accidental zeros of the one-loop expression~\eqref{eq:dmu}.}
\label{fig:dk}
\end{figure}

\paragraph*{Observables.} The form factor modifies the Higgs couplings in a way that depends on momentum.
\begin{itemize}
\item \emph{On-shell couplings.} In the decays of an on-shell Higgs boson the relative momenta are small, e.g.\ $k_E^2=m_h^2/4-m_b^2$ for $h\to b\bar b$, with $m_b$ the bottom-quark mass. All couplings are then reduced by nearly the same factor $\kappa=F_q(k_E^2)\simeq1-m_h^2/(4M^2)$, the ratio of the coupling to its SM value, almost independently of $q$. Numerically, $1-\kappa=1.6\%$, $0.39\%$ and $0.04\%$ for $M=0.5$, $1$ and $3$~TeV. Figure~\ref{fig:dk} shows the tuning as a function of this shift. $\Delta\leq10$ corresponds to $1-\kappa\gtrsim(1.7$--$3.6)\times10^{-4}$, and $\Delta\leq1$ to $1-\kappa\gtrsim(1.2$--$4.7)\times10^{-3}$.
\item \emph{Off-shell Higgs production.} The Higgs boson can also be produced far above its mass shell, in $gg\to h^*\to ZZ$ with a $ZZ$ invariant mass $m_{ZZ}$ well above $m_h$. CMS and ATLAS have found evidence for this process at a rate consistent with the SM~\cite{CMS:2022ley,ATLAS:2023dnm}. The measured ratio of the observed to the predicted rate, the off-shell signal strength $\mu_{\rm off}$, therefore limits any suppression of the Higgs couplings at high momentum. In this process the relative momentum of the two $Z$ bosons satisfies $k_E^2=m_{ZZ}^2/4-m_Z^2$, which is spacelike and grows with $m_{ZZ}$, and the $hZZ$ vertex is suppressed by $F_q(k_E^2)$. The production vertex $ggh$ lies inside the top-quark loop and is not included, which makes the estimate conservative. The rate is then suppressed by $F_q(k_E^2)^2$. Requiring it to stay within the range allowed at the 68\% confidence level (CL), $\mu_{\rm off}\geq0.5$ from the ATLAS measurement, for $m_{ZZ}=350$--$500$~GeV, where most of the signal lies, gives $M\gtrsim0.25$--$0.4$~TeV for all benchmarks (Appendix~\ref{app:offshell}). This is only an order-of-magnitude estimate. At 95\% CL the data do not exclude even $M=m_h$. A realistic composite vertex may also depend on the momentum of the off-shell Higgs, $p^2=m_{ZZ}^2$, which is set by the unknown internal dynamics, adding a further uncertainty.

\item \emph{Loop-induced processes.} The couplings $hgg$ and $h\gamma\gamma$ and di-Higgs production $gg\to hh$ proceed through top-quark (and $W$) loops, in which the nonlocal Yukawa vertex depends on the loop momentum. Coupling gluons and photons to these loops consistently requires the gauge-invariant completion of the vertices (Sec.~\ref{sec:lagrangian}), and their calculation is left for future work. The momentum-dependent trilinear Higgs vertex, which enters di-Higgs production through the triangle diagram, is given in Appendix~\ref{app:trilinear}. It grows above the SM value as $\kappa_\lambda(s)-1\simeq(s+2m_h^2)/(3M^2)$, a few percent near threshold for $M=1$~TeV. It is not by itself an observable prediction, however, since the top-quark box diagram dominates $gg\to hh$.

\item \emph{Unitarity and new states.} In the SM, Higgs exchange cancels the growth with energy, $\propto s/v^2$, of the scattering amplitudes of longitudinal $W$ bosons, which would otherwise violate unitarity~\cite{Lee:1977eg}. If the Higgs couplings are suppressed above $M$, this cancellation is incomplete, and, as in composite Higgs models, other states of the composite sector must appear near $M$ to restore it. They correspond to the timelike singularities of $F_q$ at $k^2=M^2/(q-1)$ and could be observed as heavy resonances or broad excesses in $gg\to ZZ$, $hh$ and $t\bar t$~\cite{Bellazzini:2015cgj,Shirazi:2019xjd}.
\end{itemize}
For natural values of $M$ the on-shell coupling shifts, $1-\kappa\sim10^{-4}$--$10^{-3}$, are below the precision expected at the High-Luminosity LHC~\cite{Cepeda:2019klc}, whereas the suppression by the form factor grows with momentum. The main signatures of a natural self-similar Higgs are therefore in off-shell and high-momentum Higgs processes and in the new states near $M$.

%=====================================================================
\section{Renormalization-group consequences}\label{sec:rg}
%=====================================================================
\begin{figure}[t]
\centering
\includegraphics[width=0.85\columnwidth]{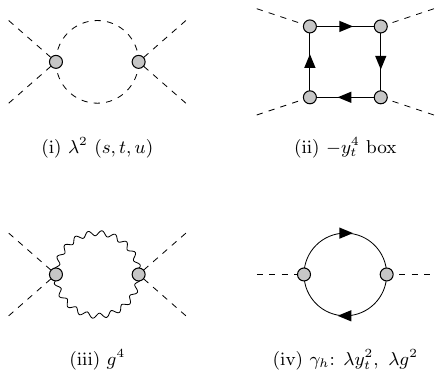}
\caption{One-loop contributions to the running of the Higgs quartic coupling, namely (i) scalar bubbles, (ii) top box, (iii) gauge loops and (iv) Higgs wave-function renormalization.}
\label{fig:beta}
\end{figure}

The form factor also modifies the scale dependence of couplings. In a single-scalar model with interaction $\lambda\phi^4/4!$, the one-loop four-point function receives only the $s$-, $t$- and $u$-channel bubble diagrams of Fig.~\ref{fig:beta}(i), since the one-loop two-point function does not depend on momentum. Each bubble carries $F_q$ at both vertices, evaluated at the relative momentum of the internal pair. In a momentum-subtraction scheme, $\lambda(\mu_R)$ is defined by the value of the four-point function at Euclidean external momentum $P_E^2=\mu_R^2$, and $\beta_\lambda=\dd\lambda/\dd\ln\mu_R$ is the Callan--Symanzik beta function in this scheme~\cite{Peskin:1995ev}. A mass-independent scheme such as $\overline{\rm MS}$, in which $\beta_\lambda$ is read off from the divergence of the bubble, cannot be used here, since with the form factor the bubble is finite. The scale $M$ makes the running mass dependent, as for the decoupling of heavy particles. In this momentum-subtraction scheme the beta function becomes (Appendix~\ref{app:beta})
\begin{equation}
\beta_\lambda\equiv\frac{\dd\lambda}{\dd\ln\mu_R}=\frac{3\lambda^2}{16\pi^2}\,R_q\!\left(\frac{\mu_R^2}{M^2}\right),
\label{eq:beta}
\end{equation}
where the suppression factor $R_q$, given in closed form in Eq.~\eqref{eq:Rq}, equals one for $F_q=1$. It is the ratio of the scale derivative of the bubble with the form factor to its SM value, i.e.\ the fraction of the SM running that survives. One has $R_q\simeq1$ for $\mu_R\ll M$ and $R_q\to0$ for $\mu_R\gg M$, where the form factor suppresses the loop contributions from momenta above $M$. For $q=4/3$, $R_q=0.98$, $0.51$, $1.3\times10^{-3}$ and $1.4\times10^{-7}$ at $\mu_R/M=0.1$, $1$, $10$ and $100$, and $R_q\propto M^4/\mu_R^4$ for $\mu_R\gg M$. The coupling thus stops running above $M$.

\begin{figure}[t]
\centering
\includegraphics[width=0.95\columnwidth]{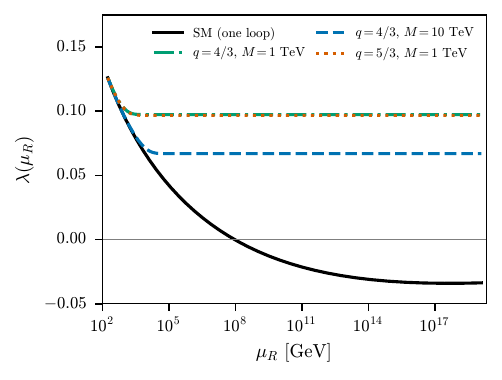}
\caption{One-loop running of the Higgs quartic coupling in the SM and with the self-similar form factor, Eq.~\eqref{eq:beta}, applied as an approximation to the full SM $\beta_\lambda$.}
\label{fig:running}
\end{figure}

In the SM, the one-loop beta function of the quartic coupling is $16\pi^2\beta_\lambda=24\lambda^2-6y_t^4+\frac38[2g^4+(g^2+g'^2)^2]+\lambda(12y_t^2-9g^2-3g'^2)$. As an approximation, since every term arises from loops attached to external Higgs legs (Fig.~\ref{fig:beta}), we weight it by $R_q$ and keep SM running for the gauge and Yukawa couplings. Starting from $\lambda(m_t)=0.126$~\cite{Buttazzo:2013uya}, the SM quartic coupling turns negative at $\mu_R\simeq10^{8}$~GeV at one loop ($\sim10^{10}$~GeV at higher orders~\cite{Buttazzo:2013uya}), whereas with the form factor it freezes at $\lambda\simeq0.097$ ($0.067$) for $M=1$~(10)~TeV, for all benchmarks (Fig.~\ref{fig:running}). Within the effective description the running of $\lambda$ thus stops near $M$. Consequently, for $M$ in the TeV range the region $\mu_R\sim10^{8}$--$10^{10}$~GeV where the SM quartic coupling turns negative is never reached. Whether the vacuum is stable at much higher field values depends on the dynamics above $M$, where the effective description breaks down, and on the UV completion. Absolute stability is not established here. Applying $R_q$ to the full SM $\beta_\lambda$, with SM running for $y_t$ and the gauge couplings, is an approximation, since these couplings would also be affected by the form factor. A term-by-term treatment, e.g.\ of the top box with four smeared vertices, and of the gauge terms, which require the gauge-invariant completion of Sec.~\ref{sec:lagrangian}, is left for future work.

%=====================================================================
\section{Confinement and the ultraviolet completion: an outlook}\label{sec:uv}
%=====================================================================
The effective description does not explain why the constituents are confined or why $M\ll M_{\rm Pl}$. Both are explained if the Higgs is a bound state of an asymptotically free ``hypercolor'' gauge theory $SU(N_{\rm HC})$ with $N_F$ fermion flavours, in the spirit of technicolor and composite Higgs models~\cite{Weinberg:1996kr,Kaplan:1983fs,Agashe:2004rs}. With hypercolor coupling $g_{\rm HC}$ defined at a high scale $\Lambda_{\rm UV}$, the confinement scale is $\Lambda_{\rm HC}=\Lambda_{\rm UV}\exp[-8\pi^2/(b_0\,g_{\rm HC}^2(\Lambda_{\rm UV}))]$, where $b_0=(11N_{\rm HC}-2N_F)/3$ is the one-loop beta-function coefficient. $\Lambda_{\rm HC}$ is exponentially smaller than $\Lambda_{\rm UV}$. For example, $\Lambda_{\rm HC}\sim1$~TeV and $\Lambda_{\rm UV}=M_{\rm Pl}$ require only $g^2_{\rm HC}(M_{\rm Pl})\simeq2.1/b_0$. The compositeness scale is set by the confinement scale, $M\sim\Lambda_{\rm HC}$, up to a model-dependent factor of order one.

In the underlying theory there is no elementary Higgs field, and an elementary field with the quantum numbers of the Higgs would have $\mathcal{Z}=0$~\cite{Weinberg:1965zz}. The effective Higgs field of Sec.~\ref{sec:lagrangian} is instead normalized canonically, with $\mathcal{Z}\simeq1$, and its compositeness is carried by the form factors. A light Higgs, $m_h\ll\Lambda_{\rm HC}$, would be natural if it is a pseudo-Nambu--Goldstone boson~\cite{Agashe:2004rs,Panico:2015jxa}. A possible origin of the self-similarity is that a confining theory that is approximately scale invariant over a range of energies, rather than a separate theory at each level, produces bound states whose structure looks the same at different resolutions, as the parton cascade of QCD does in deep inelastic scattering.

As a conjecture, the entropic index might be linked to the running gauge coupling. For QCD with $N_c$ colours and $n_f$ flavours, matching the beta function of a thermofractal effective coupling to the one-loop beta function gives $q-1=3/(11N_c-2n_f)$, i.e.\ $q\simeq1.14$ for $N_c=3$ and $n_f=6$~\cite{Deppman:2019yzp}. If the same relation held for hypercolor, $q-1=1/b_0$.

For $N_{\rm HC}=3$ with fundamental flavours, $b_0=(33-2N_F)/3$ vanishes at $N_F=16.5$, above which the theory is no longer asymptotically free. Somewhat below this value the coupling flows at low energies to a fixed value, and the theory becomes scale invariant instead of confining, with no bound states. This range of $N_F$ is the conformal window. Its lower edge, below which the theory confines as QCD does, lies at strong coupling and is not calculable in perturbation theory. The ladder approximation gives $N_F\simeq11.9$~\cite{Dietrich:2006cm}, and lattice studies and other estimates place it roughly in the range $N_F\approx9$--$13$, with significant uncertainty~\cite{Cacciapaglia:2020kgq,Lee:2020ihn}. Theories just below the edge are ``walking'', with a nearly constant coupling over a large energy range, which favours an approximately scale-invariant, self-similar bound state.

If $q-1=1/b_0$, the benchmarks $q=11/9$, $4/3$ and $5/3$ correspond to $N_F\simeq10$, $12$ and $14$ for $N_{\rm HC}=3$. The benchmark $q=5/3$ would then lie inside the conformal window, where no bound state forms, $q=4/3$ near its lower edge, and $q=11/9$ near or below it. Only $q\lesssim4/3$ would therefore allow a confined, composite Higgs. Testing this link requires a concrete model.

%=====================================================================
\section{Conclusions}\label{sec:conclusions}
%=====================================================================
We have proposed a Higgs boson with a statistically self-similar internal structure and studied its effect on the quantum corrections to the Higgs mass. The spectral representation shows that such a structure cannot soften the Higgs propagator without states of negative probability (ghosts), and that it must act instead through momentum-dependent vertex functions, as for hadronic form factors. In a superstatistical model of the self-similar structure, a Gamma-distributed size of the constituents turns a Gaussian form factor into a $q$-exponential form factor, with $q-1=2/(3N)$. Placed in the Higgs interactions, it keeps the Higgs propagator positive at leading order, up to an effect of order $m_h^2/M^2$ after symmetry breaking.

For $1<q<2$ all one-loop corrections to the Higgs mass are finite, and heavy states decouple. For the benchmarks there is little or no tuning, $\Delta\lesssim1$, for $M$ up to about $1$--$2$~TeV, depending on $q$, and the tuning stays below $10$ for $M$ up to about $3$--$5$~TeV. Natural values of $M$ imply small shifts of the on-shell Higgs couplings, $1-\kappa\simeq m_h^2/(4M^2)$, whereas the suppression by the form factor grows with momentum. The main signatures are therefore in off-shell and high-momentum Higgs processes and in new states near $M$. Current off-shell data give $M\gtrsim0.25$--$0.4$~TeV, an illustrative 68\% CL estimate. In an approximate treatment the Higgs quartic coupling stops running above $M$, and for $M$ in the TeV range the region where the SM coupling turns negative is not reached.

The hierarchy problem is thus converted into the requirement of new composite dynamics near the TeV scale, which a confining, nearly scale-invariant gauge theory could provide. Natural extensions of this work include a gauge-invariant construction of the nonlocal interactions and the study of loop-induced Higgs processes.

\appendix

\section{Spectral representation and positivity}\label{app:spectral}
\setcounter{equation}{0}
\paragraph*{K\"all\'en--Lehmann representation.} Insert a complete set of eigenstates of the full Hamiltonian between the fields in $\langle\Omega|\phi(x)\phi(y)|\Omega\rangle$. Each state is labelled by its momentum ${\bf p}$ and by the state $|n_{\bf 0}\rangle$ at rest with mass $m_n$ that it is boosted from. Lorentz invariance gives $\langle\Omega|\phi(x)|n_{\bf p}\rangle=\langle\Omega|\phi(0)|n_{\bf 0}\rangle e^{-ip\cdot x}$, and summing over ${\bf p}$ yields~\cite{Peskin:1995ev}
\begin{align}
\langle\Omega|T\phi(x)\phi(y)|\Omega\rangle&=\int_0^\infty\frac{\dd\sigma}{2\pi}\,\rho(\sigma)\,D_F(x-y;\sigma),\nonumber\\
\rho(\sigma)&=\sum_n 2\pi\,\delta(\sigma-m_n^2)\big|\langle\Omega|\phi(0)|n_{\bf 0}\rangle\big|^2\geq0,
\label{eq:KL}
\end{align}
where $D_F(x-y;\sigma)$ is the Feynman propagator of a free scalar of squared mass $\sigma$. Separating the one-particle state gives Eq.~\eqref{eq:pole}.

\paragraph*{Internal configurations and $\mathcal{Z}$.} Let the Higgs state at rest be a superposition of orthonormal internal configurations $|\xi_i\rangle$ of its constituents, all with the four-momentum of the Higgs, $|h_{\bf 0}\rangle=\sum_i c_i|\xi_i\rangle$ with $\sum_i|c_i|^2=1$. The $|\xi_i\rangle$ are not energy eigenstates, since otherwise only one of them would contribute. Writing the overlap of the field with each configuration as $A_i\equiv\langle\Omega|\phi(0)|\xi_i\rangle=\sqrt{\mathcal{Z}_i}\,e^{i\theta_i}$, with $\mathcal{Z}_i\geq0$ and phases $\theta_i$,
\begin{equation}
\mathcal{Z}=\Big|\sum_i c_iA_i\Big|^2=\sum_i|c_i|^2\mathcal{Z}_i+\sum_{i\neq j}c_ic_j^*A_iA_j^* .
\label{eq:Zfull}
\end{equation}
For a statistical ensemble of configurations, the assumption underlying the bootstrap and thermofractal descriptions, the phases are random. With $A_i=\sqrt{\mathcal{Z}_i}\,e^{i\theta_i}$ the interference term in Eq.~\eqref{eq:Zfull} is $\sum_{i\neq j}c_ic_j^*\sqrt{\mathcal{Z}_i\mathcal{Z}_j}\,e^{i(\theta_i-\theta_j)}$. For independent phases uniformly distributed on $[0,2\pi)$ one has $\langle e^{i(\theta_i-\theta_j)}\rangle=\langle e^{i\theta_i}\rangle\langle e^{-i\theta_j}\rangle=0$ for $i\neq j$, and hence
\begin{equation}
\overline{\mathcal{Z}}=\sum_i|c_i|^2\,\mathcal{Z}_i ,\qquad \min_i \mathcal{Z}_i\leq \overline{\mathcal{Z}}\leq\max_i \mathcal{Z}_i ,
\label{eq:Zi}
\end{equation}
a weighted average of the $\mathcal{Z}_i$, because the weights $|c_i|^2$ are non-negative and sum to one. Neither the ensemble average nor the pure-state value can depend on momentum, since $\mathcal{Z}$ is fixed by Lorentz invariance to be a constant.

\paragraph*{Positivity bound.} In Euclidean space the propagator is $D_E(p_E^2)=\int\dd\sigma\,\rho(\sigma)/[2\pi(p_E^2+\sigma)]$. For $\sigma\leq p_E^2$ one has $1/(p_E^2+\sigma)\geq1/(2p_E^2)$, and since $\rho\geq0$,
\begin{equation}
p_E^2\,D_E(p_E^2)\geq\frac12\int_0^{p_E^2}\frac{\dd\sigma}{2\pi}\,\rho(\sigma),
\label{eq:bound}
\end{equation}
a non-decreasing function of $p_E^2$. The propagator therefore cannot fall faster than $1/p_E^2$~\cite{Weinberg:1995mt}. For a canonically normalized field, regularized so that the equal-time commutator is canonical, $\int\dd\sigma\,\rho(\sigma)/(2\pi)=1$. This sum rule gives $0\leq \mathcal{Z}\leq1$, and $\mathcal{Z}=0$ characterizes a composite particle~\cite{Weinberg:1995mt,Weinberg:1965zz}.

\paragraph*{A $1/p^4$ propagator contains a ghost.} Consider a propagator that falls as $1/p^4$, e.g.\ $i\mathcal{Z}m^2/[k^2(k^2-m^2)]$. Partial fractions give
\begin{equation}
\frac{i\mathcal{Z}m^2}{k^2(k^2-m^2)}=\frac{i\mathcal{Z}}{k^2-m^2}-\frac{i\mathcal{Z}}{k^2},
\label{eq:ghost}
\end{equation}
i.e.\ a normal particle of mass $m$ and a massless state with residue of the wrong sign, a negative-norm ghost. The same ghost appears when the form factor is placed in the propagators (Sec.~\ref{sec:lagrangian}). For $q=2$ the propagator $iF_2(k^2)/(k^2-m^2)$ also falls as $1/k^4$, and it splits as
\begin{equation}
\frac{F_2(k^2)}{k^2-m^2}=\frac{M^2}{M^2-m^2}\left[\frac{1}{k^2-m^2}-\frac{1}{k^2-M^2}\right],
\label{eq:LWsplit}
\end{equation}
i.e.\ a normal particle of mass $m$ and a Lee--Wick partner of mass $M$ with negative residue~\cite{Lee:1969fy,Grinstein:2007mp}. More generally, any spectral function whose propagator falls faster than $1/p^2$ violates Eq.~\eqref{eq:bound} and therefore cannot be positive, assuming that the spectral function grows at most polynomially in $\sigma$. This is why the self-similar structure is introduced here through vertex functions and not through the Higgs propagator.

%---------------------------------------------------------------------
\section{Superstatistics and the $q$-exponential form factor}\label{app:superstat}
\setcounter{equation}{0}
Using $\int_0^\infty\dd\beta\,\beta^{a-1}e^{-c\beta}=\Gamma(a)/c^a$, the average of the Gaussian over the Gamma distribution~\eqref{eq:gamma} gives
\begin{align}
\int_0^\infty\!\dd\beta\,f(\beta)\,e^{-\beta k_E^2}
&=\left(\frac{a}{\langle\beta\rangle}\right)^{a}\left(\frac{a}{\langle\beta\rangle}+k_E^2\right)^{-a}\nonumber\\
&=\left(1+\frac{\langle\beta\rangle k_E^2}{a}\right)^{-a}.
\end{align}
With $q-1=1/a$ and $M^2=1/\langle\beta\rangle$ this is Eq.~\eqref{eq:Fq}. The variance of the Gamma distribution is $\langle\beta\rangle^2/a$, and hence $q-1={\rm var}(\beta)/\langle\beta\rangle^2$. In other words, the entropic index measures the relative fluctuation of the size parameter, as in the superstatistical interpretation of Tsallis distributions, where $q-1$ is the relative variance of the inverse temperature~\cite{Wilk:1999dr,Beck:2003}.

\paragraph*{Per-vertex versus correlated averaging.} In a loop with two vertices, such as the top loop, the Higgs turns into a pair at one vertex and the pair turns back into the Higgs at the other. We average the configuration independently at each vertex. This assumes that the configuration of the Higgs changes during the loop, and it is what a fixed nonlocal kernel such as $\mathcal K_q$ in Eq.~\eqref{eq:star} produces. If instead the configuration stayed frozen during the loop, both vertices would see the same $\beta$, and the relevant average would be $\langle e^{-2\beta k_E^2}\rangle=F_q(2k_E^2)$, with $\int_0^\infty F_q(2u)\,\dd u=M^2/[2(2-q)]$ instead of $\int_0^\infty F_q(u)^2\,\dd u=M^2/(3-q)$. Both integrals are finite for $1<q<2$ and differ by factors of order one. Which of the two applies depends on the internal dynamics of the composite, which are unknown.

\paragraph*{Algebraic identities.} Direct substitution shows that $F_q(u)^2=F_{\tilde q}(u)|_{M^2\to M^2/2}$ and $F_q(u/4)^2=F_{\tilde q}(u)|_{M^2\to 2M^2}$, with $\tilde q=(q+1)/2$, since both sides equal $[1+(q-1)u/M^2]^{-2n}$ and $[1+(q-1)u/(4M^2)]^{-2n}$, respectively. Every two-vertex integral therefore follows from the one-vertex result.

%---------------------------------------------------------------------
\section{Feynman rules and the one-loop self-energy}\label{app:diagrams}
\setcounter{equation}{0}
\paragraph*{Momentum rule of the star product.} For plane waves $A\propto e^{-ip_A\cdot x}$ and $B\propto e^{-ip_B\cdot x}$, ignoring the Wilson line, Eq.~\eqref{eq:star} gives
\begin{equation}
(A\star_qB)(x)=e^{-i(p_A+p_B)\cdot x}\int\!\dd^4y\,\mathcal K_q(y)\,e^{-i(p_A-p_B)\cdot y/2},
\end{equation}
i.e.\ the Fourier transform of $\mathcal K_q$ at the relative momentum $k=(p_A-p_B)/2$, which is $F_q(k_E^2)$ by construction. Since $F_q$ depends only on $k^2$, $\mathcal K_q$ is even and $A\star_qB=B\star_qA$. Writing $H=(G^+,(v+h+iG^0)/\sqrt2)^T$ in terms of four real fields $\phi_a$ ($\phi_1=v+h$), the antisymmetric cross terms cancel and $H^\dagger\star_qH=\frac12\sum_a\phi_a\star_q\phi_a$. Hence
\begin{equation}
\lambda(H^\dagger\star_qH)^2=\frac\lambda4\sum_{a,b}(\phi_a\star_q\phi_a)(\phi_b\star_q\phi_b).
\label{eq:Vreal}
\end{equation}
A constant $v$ carries zero momentum, and hence a bilinear $v\star_q h$ gives $F_q(p^2/4)$ for an $h$ of momentum $p$.

\paragraph*{Vertices.} From Eqs.~\eqref{eq:Lvertex} and~\eqref{eq:Vreal} the vertices carry their SM values multiplied by form factors. (i) The quartic and trilinear scalar couplings carry $F_q$ of each bilinear, summed over all assignments of the external fields to the bilinears (Appendix~\ref{app:trilinear}). (ii) The top Yukawa vertex is $-i(y_t/\sqrt2)F_q(k_E^2)$, with $k$ the relative momentum of $t$ and $\bar t$. (iii) The $hVV$ and $hhVV$ couplings carry $F_q$ of the relative momentum of the two gauge bosons, a working definition that is not yet gauge invariant (Sec.~\ref{sec:lagrangian}).

The diagram classes of Fig.~\ref{fig:diagrams} enter Eq.~\eqref{eq:dmu} as follows, with $\phi_1=\phi+h$ in Eq.~\eqref{eq:Vreal}.
\begin{itemize}
\item \emph{(a) Scalar seagull.} The quartic vertex contains two bilinears. In a contraction \emph{within} one bilinear, both loop fields sit in the same bilinear and the background field in the other. In a contraction \emph{across} the bilinears, each loop field sits in a different bilinear, paired with the background field. This is the picture of diagrams (A) (within) and (B) (across) of Fig.~\ref{fig:massterm}, with the vev insertions replaced by the background field and the two $h$ lines joined into the loop. Contractions within one bilinear ($h\star_qh$, and $G\star_qG$ for each Goldstone boson) carry one form factor, $F_q(k_E^2)$, with effective cutoff $M^2/(2-q)$, and $\partial m^2/\partial\phi=2\lambda\phi$ for $h$ and for each Goldstone boson. The across contraction of $h$ carries two form factors, $F_q(k_E^2/4)^2$, with effective cutoff $\int_0^\infty F_q(u/4)^2\dd u=4M^2/(3-q)$, and $\partial m^2/\partial\phi=4\lambda\phi$. The Goldstone bosons have no across contraction, since the background field points along $h$. In Eq.~\eqref{eq:CW} these give $8\pi^2v^2\delta m_h^2\supset-\lambda v^2[I_h^{\rm within}+3I_G+2I_h^{\rm across}]$, i.e.\ the SM seagull coefficient $6\lambda=4\lambda$ within ($\lambda$ from $h$ and $3\lambda$ from the three Goldstone bosons) $+\,2\lambda$ across ($h$ only). With $\lambda v^2=m_h^2/2$ this gives the first two lines of Eq.~\eqref{eq:dmu}, where the factor $F_q(p^2)=1+\mathcal O(m_h^2/M^2)$ of the within contraction at external momentum $p$ is dropped.
\item \emph{(c) Gauge seagull.} With $m_V^2\propto\phi^2$, $\partial m_V^2/\partial\phi=2m_V^2/v$. With the $hVV$ and $hhVV$ couplings carrying $F_q$ of the relative momentum of the two gauge bosons, a working definition that is not yet gauge invariant (Sec.~\ref{sec:lagrangian}), the loop carries one form factor, $F_q(k_E^2)$, with effective cutoff $M^2/(2-q)$. This gives $6m_W^2I_q(m_W^2;M^2)+3m_Z^2I_q(m_Z^2;M^2)$.
\item \emph{(g) Top loop.} The top mass $y_t\phi/\sqrt2$ comes from the smeared Yukawa vertex, and the loop, which passes through two such vertices, therefore carries two form factors, $F_q^2$, with $\partial m_t^2/\partial\phi=2m_t^2/v$. This gives $-12m_t^2I_{\tilde q}(m_t^2;M^2/2)$, with effective cutoff $M^2/(3-q)$, and hence $\delta m_h^2>0$, i.e.\ the top loop raises $m_h^2$.
\item \emph{(b), (d), (e) Bubbles.} The diagrams with two trilinear vertices carry $F_q^2$ and are finite. They give terms of order $m^2/(16\pi^2)$, which are not included in Eq.~\eqref{eq:dmu}.
\item \emph{(f) Ghosts.} In Landau gauge the ghosts do not couple to $h$.
\item \emph{(h) Tadpoles.} They are included automatically through the minimum condition $m_H^2+\delta m_H^2=-\lambda v^2$, which gives $\delta m_h^2=-2\,\delta m_H^2$.
\end{itemize}
For $F_q=1$ and a hard cutoff, every $I_j\to\Lambda^2$ up to logarithms, and Eq.~\eqref{eq:dmu} becomes $8\pi^2v^2\delta m_h^2=-3\Lambda^2(m_h^2+2m_W^2+m_Z^2-4m_t^2)$, which is Eq.~\eqref{eq:veltman}.

%---------------------------------------------------------------------
\section{Higgs mass term and self-couplings}\label{app:trilinear}
\setcounter{equation}{0}
\paragraph*{Expansion.} In unitary gauge, $H\to(0,(v+h)/\sqrt2)^T$, and the bilinear of the quartic term becomes
\begin{align}
H^\dagger\star_qH={}&\tfrac12\big[(v\star_qv)+(v\star_qh)\nonumber\\
&+(h\star_qv)+(h\star_qh)\big].
\label{eq:Hexp}
\end{align}
Hence $\lambda(H^\dagger\star_qH)(H^\dagger\star_qH)=\frac\lambda4\sum(A\star_qB)(C\star_qD)$, a sum of $16$ terms with $A,B,C,D\in\{v,h\}$. Each bilinear carries $F_q$ of the relative momentum of its pair. A constant $v$ carries zero momentum, and a bilinear of $v$ and an $h$ of momentum $p$ therefore carries $F_q(p^2/4)$ (Appendix~\ref{app:diagrams}). The terms with one Higgs field carry $F_q(0)=1$, and the minimum condition $m_H^2=-\lambda v^2$ is therefore unchanged.

\begin{figure}[t]
\centering
\includegraphics[width=0.95\columnwidth]{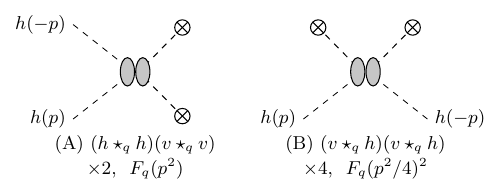}
\caption{Tree-level contributions of the nonlocal quartic term to the Higgs mass after symmetry breaking. The two ovals at the same point are the two bilinears $H^\dagger\star_qH$, the external dashed lines are the Higgs fields, with incoming momenta $p$ and $-p$, and $\otimes$ denotes a vev insertion. (A) Both Higgs fields in one bilinear, weight $F_q(p^2)$. (B) One Higgs field and one vev in each bilinear, weight $F_q(p^2/4)^2$. The multiplicities count the corresponding terms in the expansion of $(H^\dagger\star_qH)^2$ with Eq.~\eqref{eq:Hexp}.}
\label{fig:massterm}
\end{figure}

\paragraph*{Mass term.} The terms quadratic in $h$ contain two factors of $v$ and arise in two ways, shown in Fig.~\ref{fig:massterm}. (A) Both $h$ fields sit in one bilinear and both factors of $v$ in the other, $(h\star_qh)(v\star_qv)$, which occurs twice in the expansion. For a Higgs boson of momentum $p$ the two $h$ fields carry momenta $p$ and $-p$, their relative momentum is $p$, and the term carries $F_q(p^2)$. (B) One $h$ and one $v$ sit in each bilinear, which occurs four times, from the orderings $(v\star_qh)$ and $(h\star_qv)$ of each bilinear. Each bilinear then has relative momentum $p/2$, and the term carries $F_q(p^2/4)^2$. The quartic term therefore contributes
\begin{equation}
\frac{\lambda v^2}{4}\left[2F_q(p^2)+4F_q\!\left(\tfrac{p^2}{4}\right)^2\right]
\label{eq:m2quartic}
\end{equation}
to the coefficient of $h^2$ in the potential, i.e.\ $\lambda v^2[F_q(p^2)+2F_q(p^2/4)^2]$ to the squared mass. At $p=0$ this is $3\lambda v^2$, the SM value. The local term $m_H^2H^\dagger H$ adds $m_H^2=-\lambda v^2$, and hence
\begin{equation}
m^2(p^2)=\lambda v^2\left[F_q(p^2)+2F_q\!\left(\tfrac{p^2}{4}\right)^2-1\right]\simeq m_h^2\left(1+\frac{p^2}{M^2}\right),
\label{eq:m2p}
\end{equation}
with $F_q$ in its Minkowski form and $F_q(k^2)\simeq1+k^2/M^2$. At $p=0$, $m^2=2\lambda v^2=m_h^2$, as at tree level in the SM.

The pole condition $p^2=m^2(p^2)$ gives $\lambda=\lambda_{\rm SM}(1-m_h^2/M^2)$, with $\lambda_{\rm SM}=m_h^2/(2v^2)$. The residue is $\mathcal{Z}=[1-\dd m^2/\dd p^2]^{-1}_{p^2=m_h^2}\simeq1+m_h^2/M^2$. Numerically, $\lambda/\lambda_{\rm SM}=0.984$ ($0.998$) and $\mathcal{Z}=1.016$ ($1.002$) at $M=1$ ($3$)~TeV for all benchmarks. As $|p^2|\to\infty$, $F_q\to0$ and $m^2(p^2)\to m_H^2=-\lambda v^2$. The propagator therefore still falls as $1/p^2$, and the sum rule of Appendix~\ref{app:spectral} applies, which with $\mathcal{Z}>1$ gives the negative spectral weight discussed in Sec.~\ref{sec:lagrangian}.

\paragraph*{Trilinear vertex.} The trilinear vertex follows from the same counting. Its $24$ contractions are the ways of placing the three Higgs fields and one $v$ in the two bilinears, namely four positions of $v$ and $3!$ assignments of the external Higgs bosons, with incoming momenta $p_1,p_2,p_3$, to the three remaining positions in the bilinears. Each contraction has coefficient $\lambda v/4$, and for $F_q=1$ they sum to the SM vertex $-6i\lambda v$. In every contraction one Higgs boson, say $i$, sits in the $(v,h)$ bilinear, with relative momentum $p_i/2$ and factor $F_q(p_i^2/4)$, and the other two, $j$ and $k$, sit in the $(h,h)$ bilinear, with relative momentum $(p_j-p_k)/2$ and factor $F_q((p_j-p_k)^2/4)$. Each of the three choices of $i$ occurs in $8$ contractions. Therefore
\begin{equation}
\kappa_\lambda\equiv\frac{\mathcal V_{hhh}}{\mathcal V_{hhh}^{\rm SM}}=\frac13\sum_{i}F_q\!\left(\frac{p_i^2}{4}\right)F_q\!\left(\frac{(p_j-p_k)^2}{4}\right),
\label{eq:hhh}
\end{equation}
with $(i,j,k)$ a cyclic permutation of $(1,2,3)$, $\mathcal V_{hhh}$ the vertex function and $F_q$ in its Minkowski form.

\paragraph*{Di-Higgs production.} In $gg\to h^*\to hh$ an off-shell Higgs boson $h^*$, with invariant mass $\sqrt s$, turns into two on-shell Higgs bosons. Its momentum $p_1$ therefore satisfies $p_1^2=s$, and the two final-state Higgs bosons have $p_2^2=p_3^2=m_h^2$. As for the mass term, each contraction places one of the three Higgs fields together with $v$ in one bilinear and the other two together in the second bilinear. Only two situations arise.

(i) The off-shell $h^*$ sits with $v$ (one third of the contractions). The $(v,h^*)$ bilinear has relative momentum $p_1/2$, with $k^2=s/4$, which is timelike, and the form factor grows with $s$. The two final-state Higgs bosons share the other bilinear, with $k^2=m_h^2-s/4$, i.e.\ $k_E^2=s/4-m_h^2$, which is spacelike above threshold, and their form factor suppresses the vertex.

(ii) A final-state Higgs boson sits with $v$ (two thirds of the contractions). The $(v,h)$ bilinear has $k^2=m_h^2/4$ and gives only a small fixed correction. The other bilinear pairs $h^*$ with the remaining final-state Higgs. Since an outgoing particle enters with incoming momentum $-p^{\rm out}$, the relative momentum of this pair is $(p_1+p^{\rm out})/2$ rather than a difference, with $k^2=(2s+m_h^2)/4\simeq s/2$, large and timelike, which enhances the vertex.

Equation~\eqref{eq:kls} is the weighted average of situation (i), which makes up one third of the contractions, and situation (ii), which makes up two thirds,
\begin{equation}
\kappa_\lambda(s)=\frac13\Big[F_q\!\left(\tfrac s4\right)F_q\!\left(m_h^2-\tfrac s4\right)+2F_q\!\left(\tfrac{m_h^2}{4}\right)F_q\!\left(\tfrac s2+\tfrac{m_h^2}{4}\right)\Big],
\label{eq:kls}
\end{equation}
with $F_q$ in its Minkowski form. To first order in $1/M^2$, using $F_q(k^2)\simeq1+k^2/M^2$, one finds $\kappa_\lambda(s)-1\simeq(s+2m_h^2)/(3M^2)$, independently of $q$. Table~\ref{tab:kls} lists the full result of Eq.~\eqref{eq:kls}. The vertex becomes singular where $s/2+m_h^2/4=M^2/(q-1)$, i.e.\ near $\sqrt s\simeq3.0$, $2.45$ and $1.7$~TeV for $q=11/9$, $4/3$ and $5/3$ at $M=1$~TeV, and the effective description breaks down there. The momentum-dependent mass term, Eq.~\eqref{eq:m2p}, also enhances the $h^*$ propagator by about $1.7$--$2\%$ at $\sqrt s=400$--$600$~GeV for $M=1$~TeV. The top-quark box diagram, which dominates $gg\to hh$, is not included (Sec.~\ref{sec:pheno}).

\begin{table}[t]
\caption{Shift $\kappa_\lambda(s)-1$ (in \%) of the trilinear vertex in $gg\to h^*\to hh$, Eq.~\eqref{eq:kls}, at $\sqrt s=250$, $400$ and $600$~GeV.}
\label{tab:kls}
\begin{ruledtabular}
\begin{tabular}{lcccccc}
 & \multicolumn{3}{c}{$M=1$~TeV} & \multicolumn{3}{c}{$M=3$~TeV}\\
\cline{2-4}\cline{5-7}
$q$\quad$\sqrt s$ (GeV) & $250$ & $400$ & $600$ & $250$ & $400$ & $600$\\
\hline
$\to1$ & $3.2$ & $6.6$ & $14.3$ & $0.35$ & $0.71$ & $1.46$\\
$11/9$ & $3.2$ & $6.7$ & $14.7$ & $0.35$ & $0.71$ & $1.47$\\
$4/3$ & $3.2$ & $6.7$ & $14.9$ & $0.35$ & $0.71$ & $1.47$\\
$5/3$ & $3.2$ & $6.9$ & $15.5$ & $0.35$ & $0.71$ & $1.48$\\
\end{tabular}
\end{ruledtabular}
\end{table}

%---------------------------------------------------------------------
\section{The master integral}\label{app:master}
\setcounter{equation}{0}
\paragraph*{Effective cutoff.} With $n=1/(q-1)$ and $u=M^2\tau/(q-1)$,
\begin{align}
\Lambda_{\rm eff}^2&=\int_0^\infty\!\dd u\,F_q(u)=\frac{M^2}{q-1}\int_0^\infty\!\frac{\dd\tau}{(1+\tau)^{n}}\nonumber\\
&=\frac{M^2}{(q-1)(n-1)}=\frac{M^2}{2-q},
\end{align}
convergent for $n>1$, i.e.\ $q<2$. The four-dimensional measure is $\dd^4k_E/(2\pi)^4=u\,\dd u/(16\pi^2)$ after the angular integration, and $u/(u+m^2)=1-m^2/(u+m^2)$ gives the decomposition~\eqref{eq:master}.

\paragraph*{Exact form of $J_q$.} $J_q$ can be written exactly as $J_q=(1/n)\,{}_2F_1(1,n;n+1;1-\eta)$, with $\eta\equiv(q-1)m^2/M^2$ and ${}_2F_1$ the Gauss hypergeometric function, and its expansion for $\eta\to0$ gives the asymptotic form in Eq.~\eqref{eq:J}.

\paragraph*{The case $q=2$.} The monopole form factor $F_2=M^2/(M^2+u)$ makes the one-vertex integral logarithmically divergent. With a hard cutoff $\Lambda$ on $u=k_E^2\leq\Lambda^2$, partial fractions give
\begin{equation}
I_2(m^2;M^2)=\frac{M^2}{M^2-m^2}\bigg[M^2\ln\frac{\Lambda^2+M^2}{M^2}-m^2\ln\frac{\Lambda^2+m^2}{m^2}\bigg].
\label{eq:I2}
\end{equation}

\paragraph*{Heavy states.} For $M_X\gg M$ the behaviour of $I_q(M_X^2;M^2)=\int_0^\infty\dd u\,uF_q(u)/(u+M_X^2)$ depends on whether $\int^\infty u\,F_q\,\dd u$ converges, i.e.\ on whether $n>2$, which gives three cases, and in all of them the contribution decreases with $M_X$.
\begin{itemize}
\item For $1<q<3/2$ ($n>2$) the integral is dominated by $u\lesssim M^2$, and $I_q\simeq M_X^{-2}\int_0^\infty uF_q\,\dd u=M_X^{-2}[M^2/(q-1)]^2B(2,n-2)$, with $B$ the Euler Beta function, which gives the first line of Eq.~\eqref{eq:heavy}.
\item For $3/2<q<2$ ($1<n<2$) the integral is dominated by $u\sim M_X^2$, where $F_q\simeq[M^2/((q-1)u)]^n$. Using $\int_0^\infty\dd u\,u^{s-1}/(u+M_X^2)=M_X^{2(s-1)}\pi/\sin(\pi s)$ with $s=2-n$ gives the third line.
\item For $q=3/2$ ($n=2$), $uF_q\simeq4M^4/u$ for $u\gg M^2$, and the region $M^2\lesssim u\lesssim M_X^2$ gives $4M^4\ln(M_X^2/M^2)/M_X^2$. The approach to this form is slow. Numerically, the ratio is $0.88$ at $M_X/M=10^3$.
\end{itemize}
For $q=2$ with a cutoff and $M\ll M_X\ll\Lambda$, Eq.~\eqref{eq:I2} gives $I_2(M_X^2;M^2)\simeq M^2\ln(\Lambda^2/M_X^2)$ and $I_2(0;M^2)\simeq M^2\ln(\Lambda^2/M^2)$, so that $I_2(M_X^2;M^2)-I_2(0;M^2)\simeq-M^2\ln(M_X^2/M^2)$, the non-decoupling term of $\delta m_h^2$ quoted in Sec.~\ref{sec:mass}.

%---------------------------------------------------------------------
\section{Off-shell Higgs production}\label{app:offshell}
\setcounter{equation}{0}
In $gg\to h^*\to ZZ$ the two $Z$ bosons have outgoing momenta $k_1$, $k_2$ with $(k_1+k_2)^2=m_{ZZ}^2$ and $k_1^2=k_2^2=m_Z^2$. Their relative momentum $k=(k_1-k_2)/2$ satisfies $k^2=m_Z^2-m_{ZZ}^2/4$, i.e.\ $k_E^2=m_{ZZ}^2/4-m_Z^2>0$ above the $ZZ$ threshold, $m_{ZZ}>2m_Z$. The form factor of the $ggh$ vertex, which is not included (Sec.~\ref{sec:pheno}), depends on the loop momentum of the top loop and is tied to the open gauge-invariance problem of Sec.~\ref{sec:lagrangian}.
ATLAS measures $\mu_{\rm off}=1.1^{+0.7}_{-0.6}$~\cite{ATLAS:2023dnm}, whose lower edge at 68\% CL is $\mu_{\rm min}=0.5$, and the CMS measurement~\cite{CMS:2022ley} requires only $\mu_{\rm off}>0.0061$ at 95\% CL. The form factor rescales the predicted off-shell rate to $\mu_{\rm off}\simeq F_q(k_E^2)^2$, and agreement with the data therefore requires $F_q(k_E^2)^2\geq\mu_{\rm min}$, which gives
\begin{equation}
M^2\geq\frac{(q-1)\,k_E^2}{\mu_{\rm min}^{-(q-1)/2}-1},
\label{eq:offbound}
\end{equation}
or $M^2\geq2k_E^2/\ln(1/\mu_{\rm min})$ for $q\to1$. For $\mu_{\rm min}=0.5$ and $m_{ZZ}=350$ ($500$)~GeV this gives $M\geq254$ ($395$), $249$ ($388$), $246$ ($384$) and $239$ ($373$)~GeV for $q\to1$, $11/9$, $4/3$ and $5/3$. For $M=m_h$, $F_q^2\approx0.06$--$0.14$ at $m_{ZZ}=350$~GeV. Thus $M=m_h$ is disfavoured at 68\% CL, although it is not excluded at 95\% CL.

%---------------------------------------------------------------------
\section{The beta-function suppression factor}\label{app:beta}
\setcounter{equation}{0}
In the single-scalar model with interaction $\lambda\phi^4/4!$, the one-loop four-point amplitude in Euclidean space is $\Gamma_4=\lambda-(\lambda^2/2)[\mathcal B(s)+\mathcal B(t)+\mathcal B(u)]$, where $\mathcal B$ is the one-loop bubble integral of Eq.~\eqref{eq:bubble} below, with $F_q$ at both vertices. In the momentum-subtraction scheme $\lambda(\mu_R)$ is defined as $\Gamma_4$ at the symmetric Euclidean point, where each channel has $P_E^2=\mu_R^2$. Differentiating with respect to $\ln\mu_R$ gives $\beta_\lambda=-3\lambda^2\,\dd\mathcal B/\dd\ln P_E^2$ at $P_E^2=\mu_R^2$, and therefore
\begin{equation}
\beta_\lambda=\frac{3\lambda^2}{16\pi^2}\,R_q,\qquad R_q\equiv-16\pi^2\frac{\dd\mathcal B}{\dd\ln P_E^2}\bigg|_{P_E^2=\mu_R^2},
\label{eq:betaRq}
\end{equation}
which is Eq.~\eqref{eq:beta}. For $F_q=1$ the bubble behaves as $\mathcal B\simeq\ln(\Lambda^2/P_E^2)/(16\pi^2)$ with a cutoff $\Lambda$, hence $R_q=1$ and $\beta_\lambda=3\lambda^2/(16\pi^2)$, the standard result.

For massless internal lines with momenta $P_E/2\pm k_E$ and the form factor $F_q(k_E^2)$ at both vertices, the bubble is
\begin{equation}
\mathcal B(P_E^2)=\int\!\frac{\dd^4k_E}{(2\pi)^4}\,\frac{F_q(k_E^2)^2}{(k_E^2+P_E^2/4)^2-(P_E\cdot k_E)^2}.
\label{eq:bubble}
\end{equation}
Since $F_q$ depends only on $u \equiv k_E^2$, whereas the denominator depends also on the angle between $k_E$ and $P_E$, the integral splits into an integral over $u$, with the measure $u\,\mathrm{d}u/(16\pi^2)$ of Appendix~\ref{app:master}, and an average of the denominator over the directions of $k_E$. This average is exact and elementary.\footnote{With $\vartheta$ the angle between $k_E$ and $P_E$, the average is $\frac{2}{\pi}\int_0^\pi \mathrm{d}\vartheta\,\sin^2\vartheta\,\big[(u+z)^2 - 4uz\cos^2\vartheta\big]^{-1}$, where the weight $\sin^2\vartheta$ comes from the four-dimensional measure.} With $z \equiv P_E^2/4$ it equals $1/[z(u+z)]$ for $u<z$ and $1/[u(u+z)]$ for $u>z$, and the bubble reduces to a one-dimensional integral,
\begin{equation}
16\pi^2\mathcal B=\frac1z\int_0^z\frac{u\,F_q^2\,\dd u}{u+z}+\int_z^\infty\frac{F_q^2\,\dd u}{u+z}.
\end{equation}
Since $\dd\ln P_E^2=\dd\ln z$, differentiating with respect to $z$ gives $R_q$. The boundary terms cancel, and
\begin{align}
R_q&=-16\pi^2\,z\frac{\dd\mathcal B}{\dd z}\nonumber\\
&=\frac1z\int_0^z\!\frac{u\,F_q^2\,\dd u}{u+z}+\int_0^z\!\frac{u\,F_q^2\,\dd u}{(u+z)^2}+z\!\int_z^\infty\!\frac{F_q^2\,\dd u}{(u+z)^2},
\label{eq:Rq}
\end{align}
with $z=\mu_R^2/4$. For $F_q=1$ the three terms are $1-\ln2$, $\ln2-\frac12$ and $\frac12$, hence $R_q=1$, as required. For $\mu_R\gg M$ the first two terms each approach $z^{-2}\int_0^\infty uF_q^2\,\dd u$ and the third is negligible. Using $\int_0^\infty uF_q^2\,\dd u=M^4/[2(3-q)(2-q)]$, one finds
\begin{equation}
R_q\simeq\frac{32}{\mu_R^4}\int_0^\infty u\,F_q^2\,\dd u=\frac{16M^4}{(3-q)(2-q)\,\mu_R^4}.
\end{equation}
Numerical values for the benchmarks are given in Table~\ref{tab:Rq}. Inside a bubble, the two internal fields can also each share a bilinear with an external field. These contractions are not included and change $R_q$ by factors of order one.

\begin{table}[t]
\caption{Suppression factor $R_q$ of Eq.~\eqref{eq:Rq}.}
\label{tab:Rq}
\begin{ruledtabular}
\begin{tabular}{lrrrr}
$q$ & $\mu_R/M=0.1$ & $1$ & $10$ & $100$\\
\hline
$11/9$ & $0.976$ & $0.504$ & $1.1\times10^{-3}$ & $1.2\times10^{-7}$\\
$4/3$ & $0.976$ & $0.509$ & $1.3\times10^{-3}$ & $1.4\times10^{-7}$\\
$5/3$ & $0.976$ & $0.524$ & $2.6\times10^{-3}$ & $3.6\times10^{-7}$\\
\end{tabular}
\end{ruledtabular}
\end{table}

%\clearpage
\bibliography{hiself}

\end{document}